\UseRawInputEncoding
\documentclass[journal=jacsat,manuscript=article]{achemso}

\usepackage[T1]{fontenc} 
\usepackage{hyperref}
\usepackage{amsfonts,amsmath,amssymb,amsthm}
\usepackage{bm,nicefrac}
\usepackage{longtable, booktabs,array}
\usepackage{makecell}
\usepackage{float}

\def\tightlist{}
\def\boldsymbol{}
\SectionNumbersOn

\author{Yibo Li}
\affiliation{
  Center for Life Sciences, Academy for Advanced Interdisciplinary Studies, 
  Peking University, Beijing 100871, China}
\author{Jianfeng Pei}
\affiliation{
  Center for Quantitative Biology, Academy for Advanced Interdisciplinary 
  Studies, Peking University, Beijing 100871, China}
\email{jfpei@pku.edu.cn}
\author{Luhua Lai}
\affiliation{
  Center for Life Sciences, Academy for Advanced Interdisciplinary Studies, 
  Peking University, Beijing 100871, China}
\alsoaffiliation{
  Center for Quantitative Biology, Academy for Advanced Interdisciplinary 
  Studies, Peking University, Beijing 100871, China}
\altaffiliation{
  BNLMS, State Key Laboratory for Structural Chemistry of Unstable and Stable 
  Species, College of Chemistry and Molecular Engineering, Peking University, 
  Beijing 100871, China
}
\email{lhlai@pku.edu.cn}

\title[
  Learning to design drug-like molecules in three-dimensional
  space using deep generative models
]
  {Learning to design drug-like molecules in three-dimensional
   space using deep generative
   models}

\begin{document}

  \begin{abstract}
    Recently, deep generative models for molecular graphs are gaining more
and more attention in the field of \emph{de novo} drug design. A variety
of models have been developed to generate topological structures of
drug-like molecules, but explorations in generating three-dimensional
structures are still limited. Existing methods have either focused on
low molecular weight compounds without considering drug-likeness or
generate 3D structures indirectly using atom density maps. In this work,
we introduce Ligand Neural Network (L-Net), a novel graph generative
model for designing drug-like molecules with high-quality 3D structures.
L-Net directly outputs the topological and 3D structure of molecules
(including hydrogen atoms), without the need for additional atom
placement or bond order inference algorithm. The architecture of L-Net is 
specifically optimized for drug-like molecules, and a set of metrics is 
assembled to comprehensively evaluate its  performance. 
The results show that L-Net is capable of generating chemically
correct, conformationally valid molecules with high drug-likeness. Finally, to
demonstrate its potential in structure-based molecular design, we
combine L-Net with MCTS and test its ability to generate potential
inhibitors targeting ABL1 kinase.
  \end{abstract}
  \hypertarget{introduction}{%
\section{Introduction}\label{introduction}}

The search for molecules with good bioactivity and druggability is the
central task of \emph{de novo} drug discovery, which is complicated by
the enormous size and complexity of the chemical space
\cite{Bohacek.1996}. To facilitate the search for novel molecular
structures, a variety of computational \emph{de novo} molecular design
algorithms \cite{Wang.2000,Yuan.2011,Yuan.2020} have been developed
aiming to make the exploration of chemical space more efficient.
However, classical \emph{de novo} design algorithms require
expert-crafted rules to guarantee that the output molecule is practical,
and usually requires extensive tuning of parameters in the scoring
function or optimization algorithms to achieve good performance.

With the popularization of deep learning techniques, there is a growing
interest in applying deep generative models to the problem of \emph{de
novo} molecule design. Deep neural networks are highly expressive and
generalizable and can be trained in a fully data-driven manner, with
minimum requirements for expert knowledge. Early works have utilized
SMILES-based generative models for molecule
generation\cite{Segler.2017,Gomez-Bombarelli.2018}, while more recent
models have enabled the direct generation of molecular graphs using
graph convolutional neural network \cite{Li.2018,You.2018,Jin.2018}.
Those methods have demonstrated promising performance in a variety of
molecular design tasks, including generation based on
scaffold\cite{Li.2019}, pharmacophore\cite{Pogany.2018} and targets
\cite{Li.2018}.

Most of the existing works in this field have focused on generating 2D
(or topological) structures based on 2D conditions. However, 3D
information is highly important for designing molecules with high
bio-activity. For example, protein structures are routinely used in
docking studies for structure-based drug discovery (SBDD), and ligand
information can be used to build 3D-QSAR models for activity prediction.
It is therefore highly desirable to include 3D condition into deep
generative models for molecules, but explorations in this direction have
been rather limited, and there are still many problems needs to be
solved.

One way to incorporate 3D information is to condition the existing
SMILES-based generative models on 3D data, such as molecular shape
\cite{Skalic.2019} and pocket structure \cite{Skalic.2019ncg}. Those
methods have demonstrated effectiveness in ligand and structure-based
drug discovery. However, the output molecules by those models do not
contain 3D information, and an additional optimization step is usually
required to embed them into 3D space. A more desirable approach is to
directly generate 3D coordinates, as done previously by Gebauer et
al.\cite{Gebauer.2019}, Sim et al.\cite{Simm.2020toa} and Nesterov et
al.\cite{Nesterov.2020}. However, these methods mainly focused on
structurally simple molecules such as those in the QM9 dataset, and
their applicability to drug-like molecules is currently unclear. On the
other hand, more recent work by Ragoza et al.~\cite{Ragoza.2020} has
placed focus on drug-like molecules. Their model works by first
generating the atomic density map, converting it to atomic positions,
and finally converting those spatial points into molecular structures.
However, it is not an end-to-end method and requires multiple
components, both deep learning-based and rule-based, to work together to
generate the final molecule output.

In this work, we develop a novel deep learning model for the end-to-end
generation of drug-like molecules with high-quality 3D structures. The
model builds molecule iteratively by adding new atoms and bonds to the
existing structure step-by-step and outputs results with full 3D
coordinates, without the need for additional processing steps. To
achieve good performance in generating conformationally valid molecules,
a new network architecture, named L-Net (which stands for ``ligand
neural network''), is proposed. We design the network to be covariant
against rotation and translation, and domain knowledge is incorporated
into the network to improve the chemical validity of the generated
molecules. We show that our proposed method is capable of generating
chemically correct, conformationally valid, and highly drug-like
molecules. Finally, in order to demonstrate its potential application in
SBDD, we combine the model with Monte Carlo tree search (MCTS), a widely
applied technique in reinforcement learning, and use it to optimize
molecules targeting ABL1. To our knowledge, this work presents the first
attempt to apply an autoregressive 3D graph generative model to an SBDD
problem. The major contributions of this paper are summarized as
follows:

\begin{itemize}
\tightlist
\item
  We propose a new 3D molecule generative model called L-Net that is
  specifically designed to generate drug-like molecules.
\item
  We assemble a set of evaluation metrics to comprehensively measure the
  performance of the proposed model. Compared with previous benchmarks
  for 2D models, the proposed metrics emphasize the quality of 3D
  structures.
\item
  As a proof of concept, we combine L-Net with MCTS to design potential
  inhibitors against ABL1.
\end{itemize}

  \hypertarget{methods}{%
\section{Methods}\label{methods}}

\hypertarget{the-molecule-generation-process}{%
\subsection{The molecule generation
process}\label{the-molecule-generation-process}}

\begin{figure}[!tbh]
\centering
\includegraphics[width=\linewidth]{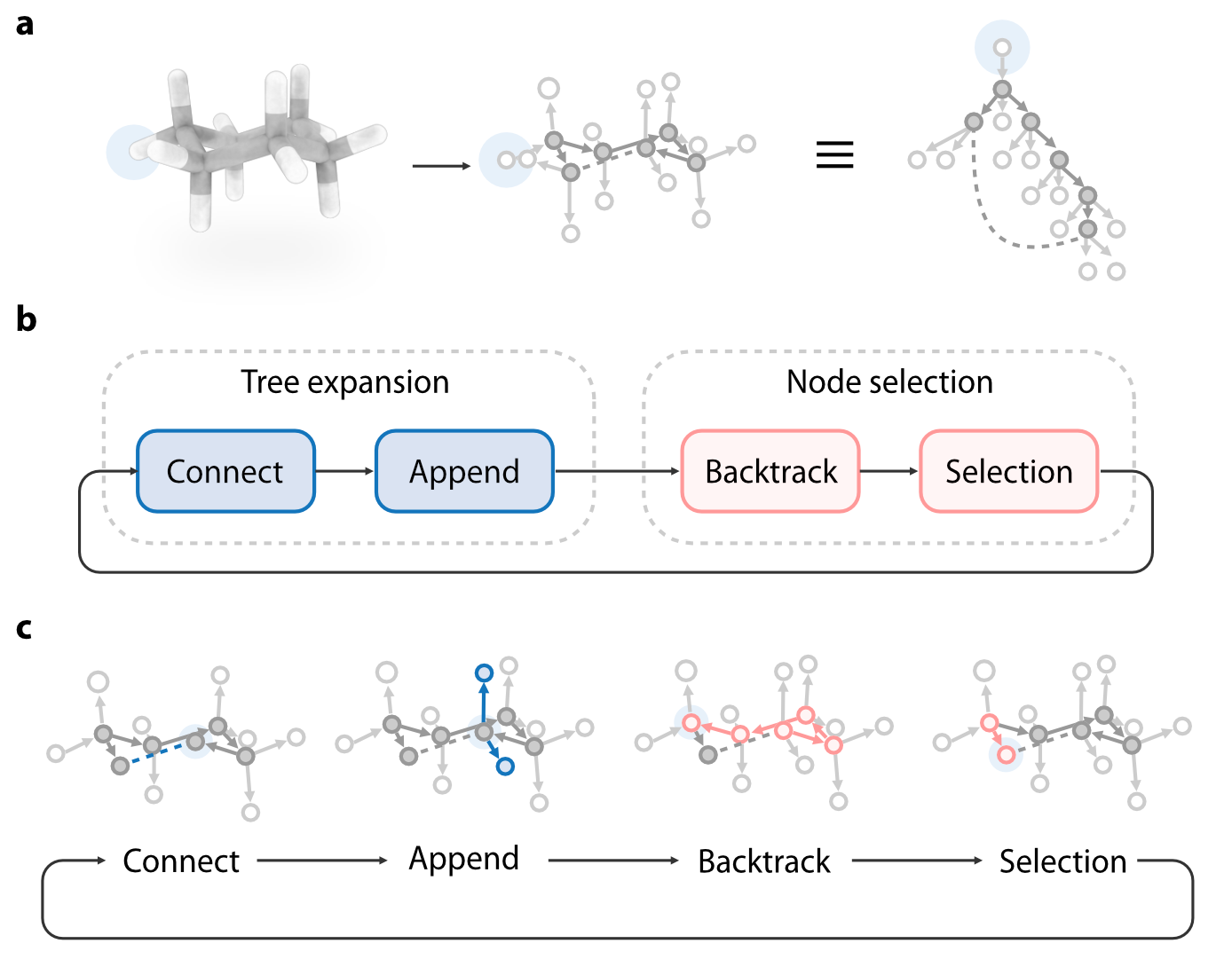}
\caption{The molecule generation process, using cyclohexane as an
example: \textbf{a.} The model generates a molecule by generating its
spanning tree in a depth-first manner. The root of the tree is
highlighted in light blue; \textbf{b.} The model builds the spanning
tree iteratively, and two sub-steps are performed at each iteration:
tree expansion and node selection. At the tree expansion step, the model
connects the currently focused atom with another atom (the ``connect''
operations), or add new atoms to the focused atom (the ``append''
operations). At the node selection step, the model backtracks the
spanning tree to find the next focus atom (the ``backtrack'' and
``selection'' operations); \textbf{d.} An example showing a few steps
during the generation of cyclohexane. Blue atoms or bonds indicates
newly added structures during the tree expansion step. Pink arrows
indicate the process of finding the next focused atom. The focused atom
at each step is highlighted using a light blue
circle.\label{fig:gen-proc}}
\end{figure}

The task of our generative model is to produce molecular graphs
\(G=(V, E, A, B, X)\) , where \(V\) is the set of nodes (atoms), \(E\)
is the set of edges (bonds), \(A=\{a_v\}_{v\in V}\) are the atom type
labels, \(B=\{b_{uv}\}_{(u, v) \in E}\) are bond type labels, and
\(X = \{\mathbf{x}_v\}_{v \in V}\) are the 3D positions of each atoms.
Note that in theory, the bond order can be inferred directly from
distances between atoms, as done in several previous works
\cite{Gebauer.2019,Ragoza.2020}, but existing bond type assignment
algorithms are generally sensitive to errors, even small ones that can
be later corrected. To make our model more robust, we explicitly output
the bond types \(B\).

Our proposed model generates the graph in a step-by-step manner. More
specifically, the model generates a molecular graph by iteratively
building its spanning tree. A spanning tree of \(G\) is a tree structure
that contains all nodes in \(G\) (see Figure
\ref{fig:gen-proc}\textbf{a}). Tree-based structures are much simpler
than general unconditional graphs, and its generation is more
straightforward using a depth-first approach. At each iteration, the
following two steps are performed to build the spanning tree (Figure
\ref{fig:gen-proc}\textbf{b}):

\begin{itemize}
\tightlist
\item
  Node selection: The model selects a ``focus atom'' from the set of
  suitable atoms that have already been generated. An atom is suitable
  for becoming a focused atom if it has unfilled valences (as an
  example, see the focused atom in Figure \ref{fig:gen-proc}\textbf{c});
\item
  Tree expansion: The model performs edits around the focused atom, by
  either adding new atoms to it (the ``append'' operation) or by
  connecting it with another existing atom (the ``connect'' operation).
\end{itemize}

During the ``node selection'' step, the model searches through the
spanning tree to find the next focused atom:

\begin{itemize}
\tightlist
\item
  If the currently focused atom has a child atom whose valence has not
  been filled, the model will select that atom to be the next focus. If
  multiple such children exist, a ranking is performed and the
  highest-ranking child is selected;
\item
  If no such child exists for the current focus, a ``backtracking''
  operation is performed to find an ancestor who has such children. And
  that child is then selected as the next focus.
\end{itemize}

This process terminates when there are no atoms suitable for becoming
the ``focused atom'', that is, the valences of all atoms have been
filled. During the generation process, there are a variety of decisions
that needs to be made by the model:

\begin{itemize}
\tightlist
\item
  During the ``connect'' operation, the model needs to decide which atom
  to connect with, using what type of bond;
\item
  During the ``append'' operation, the model needs to decide how many
  atoms should be added to the graph, their atom type, 3D location, and
  the type of bonds connecting them to the focused atom.
\item
  The model also needs to output the ranking of each atom, which will be
  used in the node selection step.
\end{itemize}

Those decisions are all made using a neural network with a novel
architecture we called L-Net. L-Net is composed of two parts: The first
part is a state encoder, which maps the intermediate molecular structure
\(G_i\) at step \(i\) into a continuous representation
\(\mathbf{h}_i = f_\boldsymbol{\theta}(G_i)\). The second part is a
policy network, which assign a probability value to each available
action based on the current state
\(p_\boldsymbol{\theta}(a|\mathbf{h}_i)\). The architecture of L-Net is
explained in detail from Section
\ref{the-architecture-of-the-state-encoder} though
\ref{pooling-and-unpooling-operations-in-graph-u-net}. To make the
network capable of generating drug-like molecules, we construct a
drug-like subset of the ChEMBL dataset \cite{Mendez.2018} and created an
``expert trajectory'' for generating each molecule in the dataset. L-Net
is then trained by imitating those trajectories. Data collection and
preprocessing workflows are given in Section
\ref{data-collection-and-preprocessing}, and the training details are
given in Section \ref{model-training}. Finally, to validate the model's
performance, we designed a set of evaluation metrics which are discussed
in Section \ref{evaluation}.

  \hypertarget{the-architecture-of-the-state-encoder}{%
\subsection{The architecture of the state
encoder}\label{the-architecture-of-the-state-encoder}}

At iteration \(i\), the state encoder of L-Net is responsible for
mapping the current molecular graph \(G_i\) to continuous
representations
\(h_i = (\mathbf{h}_{i,\mathrm{g}}, \{\mathbf{h}_{i,v}\}_{v \in V_i})=f_\boldsymbol{\theta}(G_i)\),
where \(\mathbf{h}_{i,\mathrm{g}}\) is the graph level representation,
and \(\{\mathbf{h}_{i,v}\}_{v \in V_i}\) are atom level representations.
The architecture of \(f_\boldsymbol{\theta}\) is shown in Figure
\ref{fig:architecture}. The network adopts a U-net
structure\cite{Ronneberger.2015}. The input is first fed into an
embedding layer to create the input representation for atoms and bonds.
It is then passed into the U-net, which is built from convolutional
layers, pooling layers, and unpooling layers. The convolutional layers
adopt the architecture of MPNN\cite{Gilmer.2017}, and are organized into
DenseNet blocks \cite{Huang.2016vip} to improve the performance. Pooling
layers and unpooling layers use a node clustering method that is
specifically designed for this use-case. The results are collected and
fed to the policy network.

The following sections are devoted to giving detailed explanations of
the individual components of the state encoder. We first describe the
embedding layer in Section \ref{the-embedding-layers}, and then graph
convolution layer in Section \ref{graph-convolutional-layers}. Pooling
and unpooling layers, as well as the node clustering algorithm, are
discussed in Section
\ref{pooling-and-unpooling-operations-in-graph-u-net}.

  \hypertarget{the-embedding-layers}{%
\subsection{The embedding layers}\label{the-embedding-layers}}

\hypertarget{embeddings-of-atom-and-bond-types}{%
\subsubsection{Embeddings of atom and bond
types}\label{embeddings-of-atom-and-bond-types}}

The embeddings of atom and bond types are created by indexing through a
trainable lookup table. The dimensionality of those embeddings is 2.
Note that the atom type of a node \(v\) is defined by a tuple of three
variables \(a_v = (a_v^e, a_v^c, a_v^f)\): the element type \(a_v^e\),
formal charge \(a_v^c\), and whether the atom is the focused atom
\(a_v^f\). Each of those variables is embedded separately and then
concatenated together. We also add ``temporal encodings'' to each atom
to record the time that atom is added to the graph, similar to what is
done by Vaswani et al \cite{Vaswani.2017}:

\begin{figure}[!tbh]
    \centering
    \includegraphics[width=\linewidth]{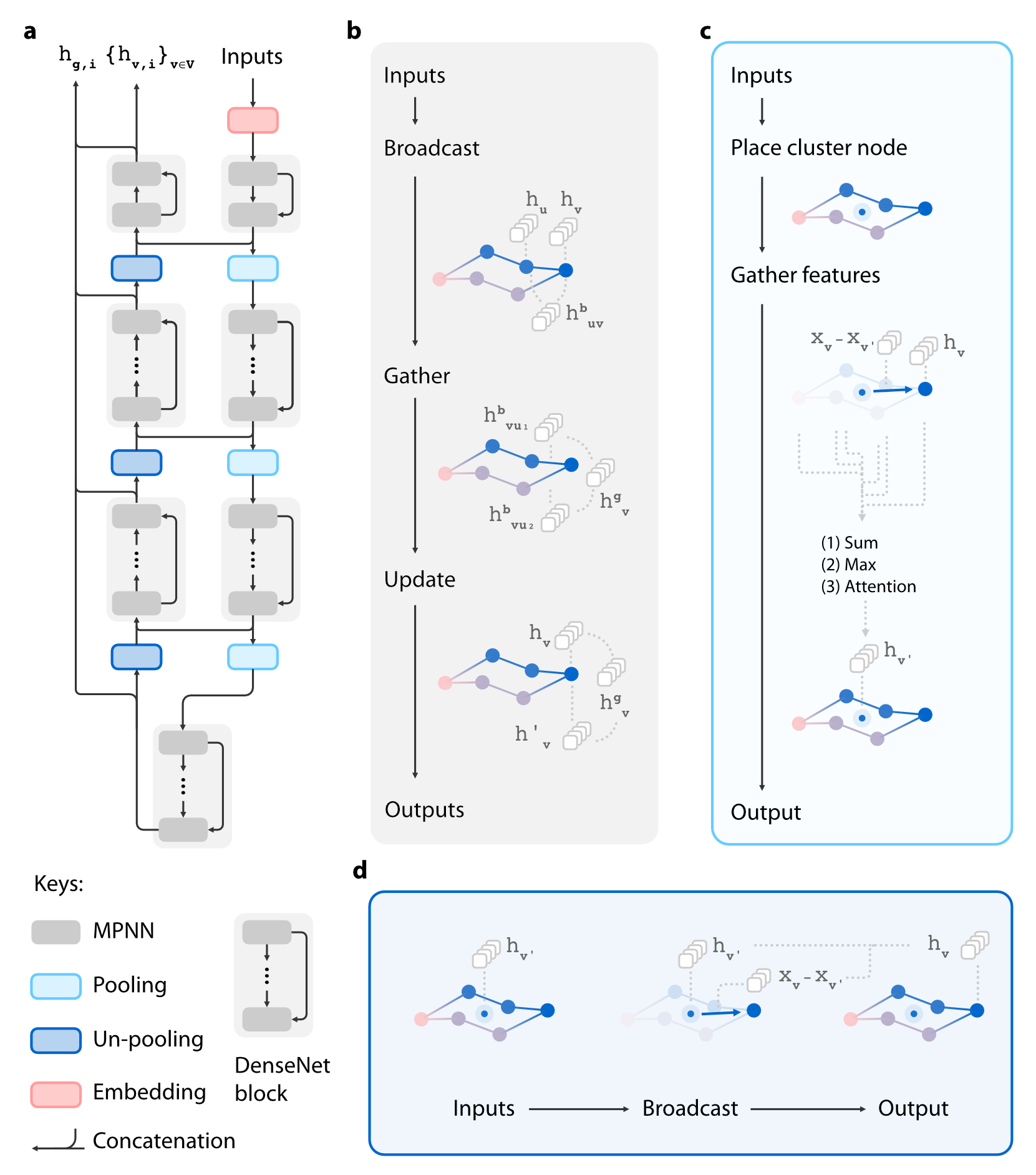}
    \caption{An overview of the architecture for the state encoder.
    \textbf{a.} The overall architecture of the state encoder; \textbf{b.}
    The architecture of each graph convolutional layer; \textbf{c.} The
    architecture of each pooling layer; \textbf{c.} The architecture of each
    unpooling layer.\label{fig:architecture}}
    \end{figure}

\[
\mathbf{h}_v^\mathrm{sin}=[\sin\frac{t_v}{T^{2l/d_\mathrm{model}}}]_{l=1}^{d_\mathrm{model}/2}
\]

\[
\mathbf{h}_v^\mathrm{cos}=[\cos\frac{t_v}{T^{2l/d_\mathrm{model}}}]_{l=1}^{d_\mathrm{model}/2}
\]

Where \(t_v\) is the step when the atom \(v\) is added to the graph,
\(d_\mathrm{model}\) is the size of the temporal embedding, \(l\) is the
indices of the temporal embedding, \(T\) determines the maximum
wavelength of the sine and cosine embedding function, \([\cdot;\cdot]\)
means the vertical concatenation of vectors. In this work,
\(d_\mathrm{model}=40\), \(T=30\). Those vectors are then projected into
a vector of size 3 with a linear layer without activation.

\hypertarget{local-coordinate-system-and-rotational-covariance}{%
\paragraph{Local coordinate system and rotational
covariance}\label{local-coordinate-system-and-rotational-covariance}}

\begin{figure}[!tbh]
\centering
\includegraphics[width=\linewidth]{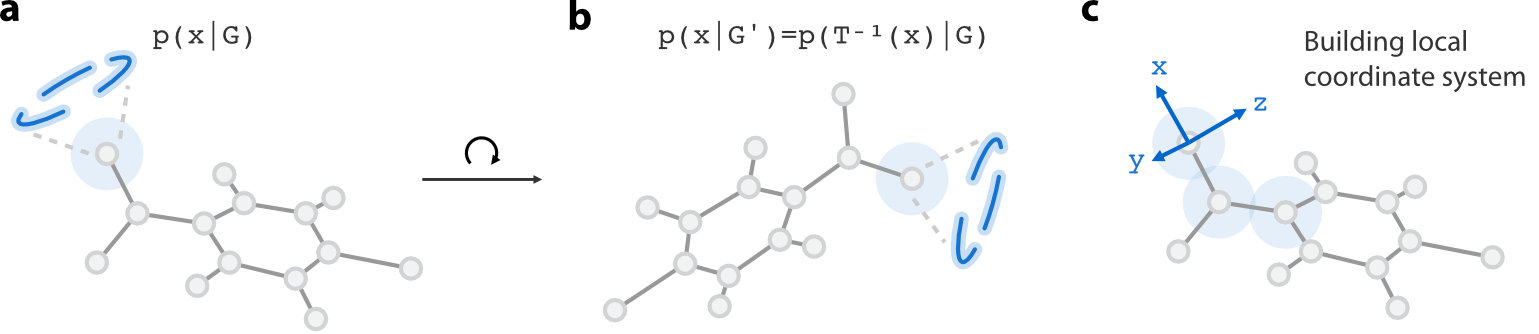}
\caption{Enforcing rotational covariance using a local coordinate
system. An ideal 3D molecule generative model should be rotationally
equivariant. If the existing structure is rotated by a certain amount
(\textbf{a}), the atom placement policy \(p(x|G)\) should also be
rotated equivalently (\textbf{b}). We solve this problem by introducing
a local coordinate system in the focused atom (\textbf{c}). Since this
coordinate system is rotationally covariant, so will the probability
distribution defined in the coordinate system.\label{fig:covariance}}
\end{figure}

Besides type information, the position of each atom should also be
included in the input. Ideally, this should be done in a rotationally
and translationally covariant way. More specifically, our model
parametrize a probability distribution (called the ``policy'') in 3D
space \(p(\mathbf{x}|G)\) to indicate where the new atom should be
placed (Figure \ref{fig:covariance}\textbf{a}). If we rotate the
existing structure \(G\) by a certain amount, the probability
distribution \(p(\mathbf{x}|G)\) should also be rotated by the same
amount (Figure \ref{fig:covariance}\textbf{b}). Mathematically, this is
expressed by the equation
\(p(\mathbf{x}|T(G))=p(T^{-1}(\mathbf{x})|G)\), where \(T\) is an
spatial operation from the special Euclidean group \(SE(3)\).

Enforcing rotational covariance into the network is a non-trivial task.
Previous works have developed specialized network architectures, such as
tensor field network \cite{Thomas.2018} or Cormorant
\cite{Anderson.2019}, to ensure rotational covariance. Those works are
theoretically elegant but are difficult to implement, and their
expressiveness may be restricted in some highly symmetrical cases.

Here, we adopt a much simpler yet effective approach, by creating a
local coordinate system at the current focus using its neighbor atoms
(Figure \ref{fig:covariance}\textbf{c}). We then express the
distribution in the local coordinate system
\(q(\mathbf{x}|G) = p(T(\mathbf{x})|T(G))\), where \(T\) indicates the
transformation from global to local coordinate. It is easy to verify
that \(q\) is rotationally covariant, even if \(p\) is not, since the
local coordinate system is covariant against rotation. Similar methods
have been previously used to construct the 3D representation of proteins
\cite{ingraham2019learning}, but to our knowledge, we are the first to
use this technique in the generative model of small molecules.

Under this framework, the 3D information feed into the neural network
are all under the local coordinate system:

\[
\tilde{\mathbf{x}}_v = M(\mathbf{x}_v-\mathbf{x}_{v'})
\]

\[
\tilde{\mathbf{x}}_{uv} = \tilde{\mathbf{x}}_u - \tilde{\mathbf{x}}_v
\]

Where \(\tilde{\mathbf{x}}_v\) is the position feature of atom \(v\),
\(\tilde{\mathbf{x}}_{uv}\) is the position feature of bond \(uv\), and $M$
is the matrix for transforming into the local coordinate system.
Those features are projected into a vector of length 8 using a linear
layer with no activation.

  \hypertarget{graph-convolutional-layers}{%
\subsection{Graph convolutional
layers}\label{graph-convolutional-layers}}

The major components of the state encoder are graph convolutional (GC)
layers. The GC architecture used in this work is similar to that used
before \cite{Li.2019}, with broadcast, gathering and update operations
parametrized using linear layers with elu activation function (as shown
in Figure \ref{fig:architecture}\textbf{b}). The only difference lies in
the gathering operation. Besides summation and maximization, we add
attention as an additional reduction method to improve the
expressiveness of the model. Also similar to the previous work, we add
``virtual'' bonds to the graph to increase the size of receptive fields
for each GC layer.

The GC layers are organized into multiple DenseNet blocks (as shown in
Figure \ref{fig:architecture}). DenseNet is a type of network
architecture that aims to increase the performance scalability for
deeper networks by introducing short connections between any two layers
\cite{Huang.2016vip}. There are three major hyper-parameter for
DenseNet: the growth rate, the bottleneck size, and the network depth.
We experiment with three configurations of DenseNet architectures:

\textbf{The standard configuration}, with bottleneck size of 94, growth
rate of 24, and the depths of DenseNet blocks (in the order of dataflow
in U-Net) {[}2, 4, 6, 8, 6, 4, 2{]};

\textbf{The shallow DenseNet}, with the same bottleneck size and growth
rate as the basic configuration, and change the depth of DenseNet blocks
to {[}2, 2, 4, 6, 4, 2, 2{]};

\textbf{The narrow DenseNet}, with the depth of each DenseNet block the
same as the basic configuration, and the bottleneck size and growth rate
reduced to half.

We will show (in Section \ref{results-and-discussion}) that reducing the
depth or width of the DenseNet blocks will both hurt the model's
performance. Since adding more layers or depth will increase the
computational burden, we suggest the use of the ``standard''
configuration for future adoption of the model.

  \hypertarget{pooling-and-unpooling-operations-in-graph-u-net}{%
\subsection{Pooling and unpooling operations in graph
U-net}\label{pooling-and-unpooling-operations-in-graph-u-net}}

U-nets \cite{Ronneberger.2015} have enjoyed great success in
image-related pixel-wise prediction tasks. It can achieve a high
receptive field size with fewer layers, while significantly reduced
memory consumption during training. The major problem for applying U-net
in graph generation is that, unlike images and 3D voxels, there are no
canonical ways to perform pooling and unpooling on graphs
\cite{Gao.2019dn}. In order to perform pooling and unpooling on
molecular graphs, we designed a custom clustering scheme:

\begin{itemize}
\tightlist
\item
  In the first level of clustering, atoms with one valence, such as
  hydrogens, halogen, and oxygens in carbonyl groups, are collapsed to
  their neighbor atoms. For most molecules, almost half of their atoms
  are hydrogen, consuming a significant amount of GPU memory. This level
  of clustering enables us to include hydrogens into the generation
  process in an efficient way, by compressing the information of
  hydrogens into its neighboring heavy atoms.
\item
  In the second level of clustering, molecules are fragmented into ring
  assemblies and chains. This method is previously used to define
  molecule scaffold\cite{Bemis.1996} and to organize scaffold datasets
  \cite{Wilkens.2005}. After fragmentation, atoms in the same ring
  assembly or chain are clustered together.
\item
  In the final level of clustering, all nodes are collapsed into a
  single graph-level master node.
\end{itemize}

\begin{figure}[!tbh]
\centering
\includegraphics{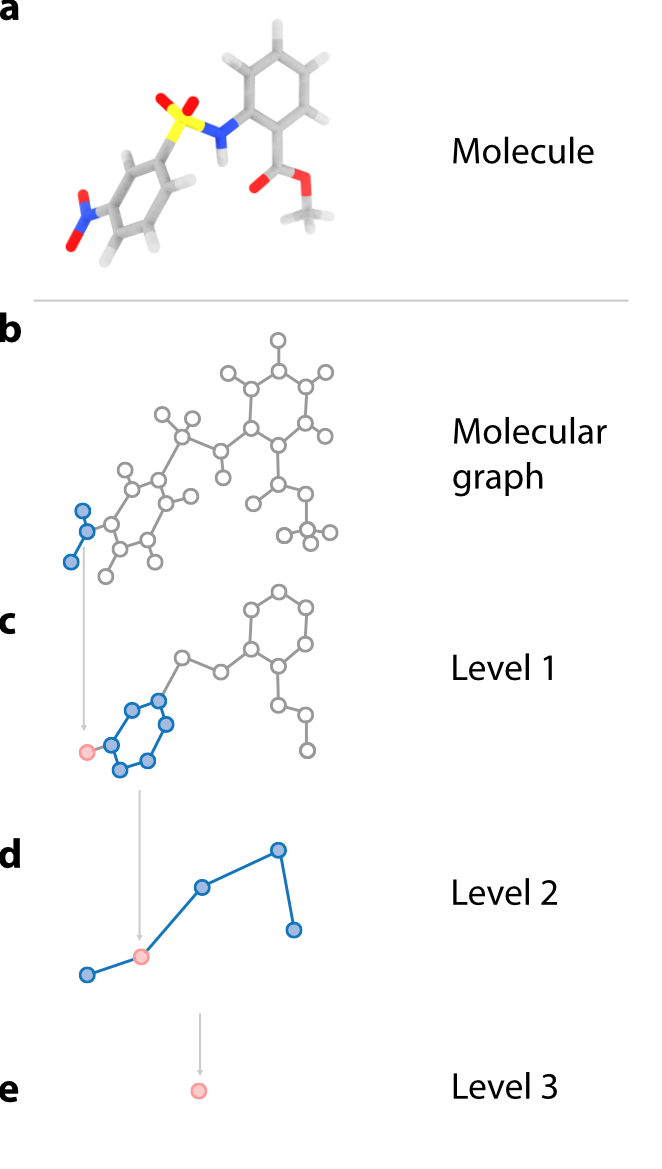}
\caption{A custom three-level node clustering scheme for pooling and
unpooling operations in molecular graphs.\label{fig:clustering}}
\end{figure}

A visual demonstration of this scheme is given in Figure
\ref{fig:clustering}. After the clustering method is be defined, the
pooling and unpooling operations can subsequently defined, as shown in
Figure \ref{fig:architecture}\textbf{c,d}. For pooling layers, we first
need to define the 3D placement of the cluster node:

\[
\Delta \mathbf{x}_v = \mathrm{MLP}_\mathrm{pooling}^\mathrm{displace}(\mathbf{h}_v)
\]

\[
\mathbf{x}_{v'} = \frac{1}{|C|} \sum_{v \in C}{(\mathbf{x}_{v} + \Delta \mathbf{x}_{v})}
\]

Where \(v'\) is the cluster node, \(C\) is the set of nodes clustered
\(v'\), and \(\mathrm{MLP}_\mathrm{Pooling}^\mathrm{Displace}\) is a
linear layer with elu activation. In other words, we place \(v'\) at the
geometric center of the nodes to be clustered and displace it with a
small deviation (initialized as zeros) calculated using a neural
network. After determining the 3D location of \(v'\), the features after
pooling is calculated as:

\[
\mathbf{h}_{v'}^\mathrm{sum} = \sum_{v \in C}\mathrm{MLP}_\mathrm{pooling}^\mathrm{sum} ([\mathbf{h}_v, \mathbf{x}_v - \mathbf{x}_{v'}])
\]

\[
\mathbf{h}_{v'}^\mathrm{max} = \max_{v \in C}\mathrm{MLP}_\mathrm{pooling}^\mathrm{max}([\mathbf{h}_v, \mathbf{x}_v - \mathbf{x}_{v'}])
\]

\[
\mathbf{h}_v^\mathrm{att}, a_v^\mathrm{att} = \mathrm{MLP}_\textrm{pooling}^\textrm{att}([\mathbf{h}_v, \mathbf{x}_v - \mathbf{x}_{v'}])
\]

\[
\mathbf{h}_{v'}^\mathrm{att} = \frac{\sum_{v \in C} \exp{a_u^\mathrm{att}} \mathbf{h}_v^\mathrm{att}}{\sum_{v \in C} \exp{a_u^\mathrm{att}} }
\]

\[
\mathbf{h}_{v'} = [
\mathbf{h}_{v'}^\mathrm{sum},
\mathbf{h}_{v'}^\mathrm{max},
\mathbf{h}_{v'}^\mathrm{att}]
\]

This may seem complicated first, but it is essentially the process of
gathering information using multiple reduction methods, including sum,
maximization, and attention, and concatenate the result together. The
unpooling layers have a much simpler architecture:

\[
\mathbf{h}_v = \mathrm{MLP}_\textrm{unpooling}([\mathbf{h}_{v'}, \mathbf{x}_{v} - \mathbf{x}_{v'}])
\]

Which uses a single fully connected layer to broadcast information to
the nodes belonging to the cluster.

  \hypertarget{the-policy-network}{%
\subsection{The policy network}\label{the-policy-network}}

After creating a continuous representation of the current state \(G_i\)
using the state encoder
\(h_i = (\mathbf{h}_{i, g}, \{\mathbf{h}_{i, v}\}_{v \in V_i})=f_\boldsymbol{\theta}(G_i)\),
the policy network is used to decide what action should be carried out.
Recall that there are three types of decision the policy network need to
make:

\begin{itemize}
\tightlist
\item
  The type and position of new atoms during the ``append'' operation;
\item
  The atom to be connected and the type of connecting bond during the
  ``connect'' operation;
\item
  The rank of the new atoms to be added.
\end{itemize}

We denote the policies for each decision as
\(p_\boldsymbol{\theta}^\mathrm{append}\),
\(p_\boldsymbol{\theta}^\mathrm{connect}\) and
\(p_\boldsymbol{\theta}^\mathrm{rank}\).

\hypertarget{decision-making-during-the-append-operation}{%
\paragraph{Decision making during the ``append''
operation}\label{decision-making-during-the-append-operation}}

\begin{figure}[!tbh]
\centering
\includegraphics[width=\linewidth]{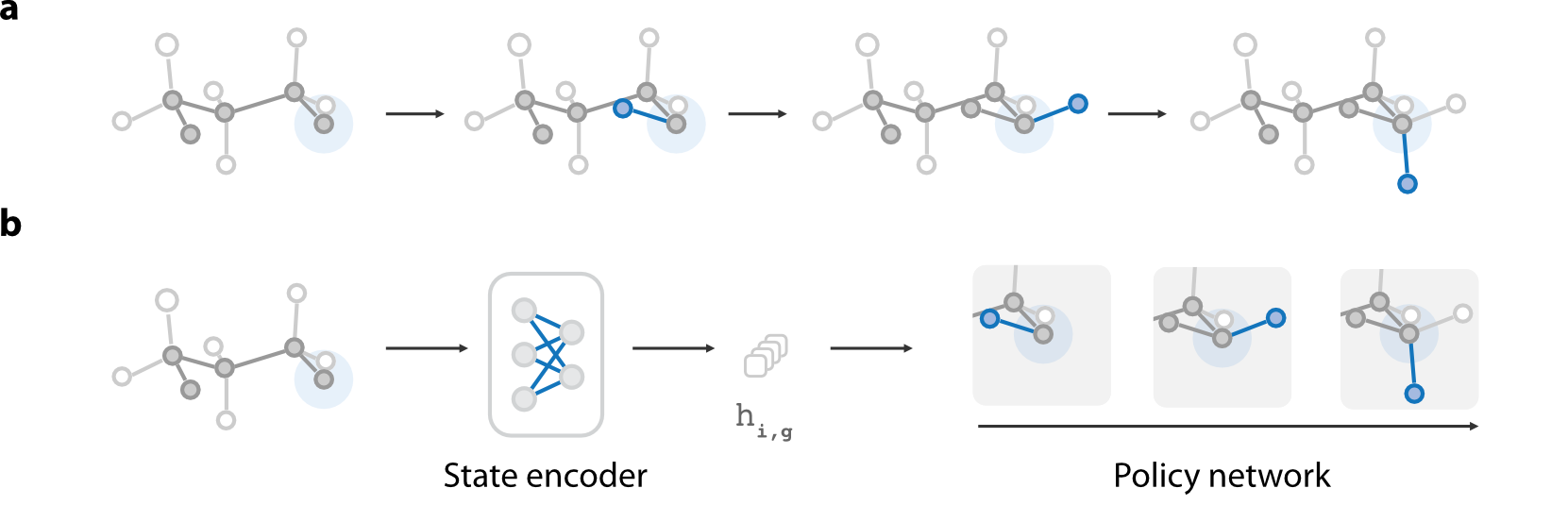}
\caption{Different from most autoregressive models in 3D molecule
generation (\textbf{a}), our proposed method generates all atoms
connected to the focused atom as a group (\textbf{b}).\label{fig:group}}
\end{figure}

During the ``append'' operation, one or more atoms are created and added
to the focused atom \(v'\). We represent a newly created atom as the
tuple \(v^*=(a, b, \mathbf{x})\), where \(a\) is the atom type, \(b\) is
the bond type used to connect the new atom with the focused atom,
\(\mathbf{x}=(r, \theta, \phi)\) is the spherical coordinate of this new
atom in the local coordinate system (described in Section
\ref{the-embedding-layers}). The policy network for the append action
can be written as: \[
p_\boldsymbol{\theta}(v^*_1,..., v^*_m|G_i) = p_\boldsymbol{\theta}(a_1, b_1, r_1, \theta_1, \phi_1, ..., a_m, b_m, r_m, \theta_m, \phi_m|G_i)
\] Where \(m\) is the number of new atoms to add. Compared to most
previous autoregressive models for 3D molecules
\cite{Gebauer.2019,Simm.2020toa}, our proposed method generates all
atoms connected to \(v^*\) in a single iteration (Figure
\ref{fig:group}). This has two major advantages. First, it can save
computational resources (since the state encoder is run only once for
each step). Secondly, since the positions of neighboring atoms are
highly correlated, generating them together can potentially improve the
model's performance.

We incorporate several prior knowledge about drug-like molecules. First,
most atoms in organic drug-like molecules have a valence less or equal
to 4. That is, at most three atoms can be added to a focused atom at one
time:

\[
p_\boldsymbol{\theta}(v^*_1,v^*_2,v^*_3|G_i)
\]

When less than three new atoms are added, we add ``null'' atoms to fill
the blanks. This makes the problem simpler since we are now dealing with
a fixed number of random variables.

Another prior knowledge we incorporate is that there are only three
types of allowed local geometry for most drug-like molecules: sp
(linear), sp2 (planar), and sp3 (tetrahedral). We ask the model to first
generate the type of the local geometry (\(h\)), and then the position
of new atoms:

\[
p_\boldsymbol{\theta}(v^*_1,v^*_2,v^*_3|h, G_i)p_\boldsymbol{\theta}(h|G_i)
\]

Empirically, we find that this can help the model to better learn the
local geometry. Now, we still need to find a way to factorize
\(p_\boldsymbol{\theta}(v^*_1,v^*_2,v^*_3|h, G_i)\), which contains (2 +
3) * 3 = 15 random variables. A natural choice is to factorize it into
an autoregressive model:

\[
p(v^*_1,v^*_2,v^*_3)=
p(v^*_3|v^*_1,v^*_2)
p(v^*_2|v^*_1)
p(v^*_1)\\
\]

\[
p(v^*_i|\cdot) = p(\phi_i|a_i, b_i, r_i, \theta_i, \cdot)
p(\theta_i|a_i, b_i, r_i, \cdot)
p(r_i|a_i, b_i, \cdot)
p(a_i, b_i, \cdot)
\]

\begin{figure}[!tbh]
\centering
\includegraphics[width=\linewidth]{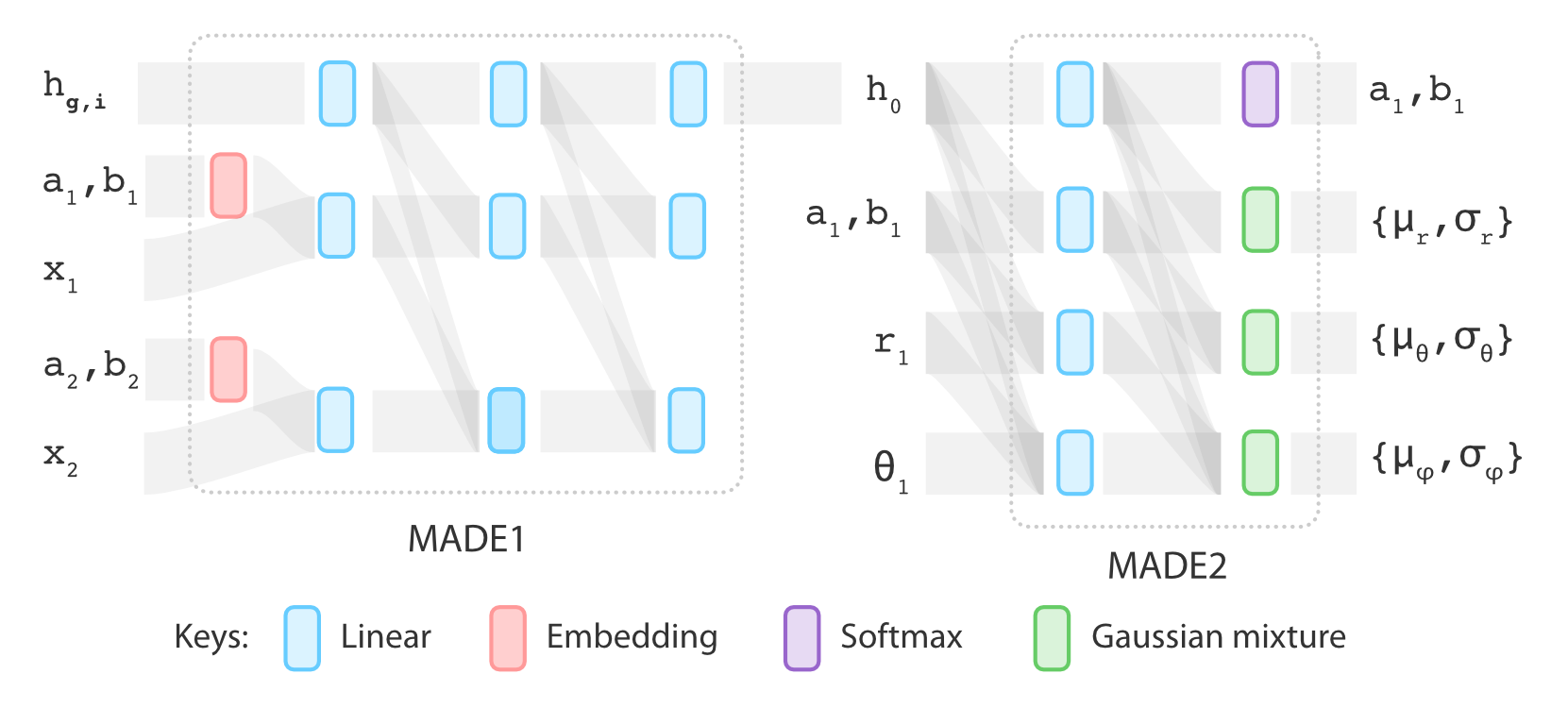}
\caption{The architecture of MADE blocks. Output sizes of linear layers
in the MADE1 block are 128, output sizes of linear layers in the MADE2
block (except that used in the Gaussian mixture layer) are 64. The
embedding table used here is the same as that used in the state encoder.
The activation function used for outputing standard deviation of \(r\),
\(\theta\) and \(\phi\) is softplus.\label{fig:made}}
\end{figure}

The conditions \(G_i\) and \(h\) are omitted for simplicity.
\(p_\boldsymbol{\theta}(a_i, b_i|\cdot)\) is a categorical distribution
of atom and bond types. We apply prior knowledge about the allowed
valence for the focused atom to create a mask for
\(p_\boldsymbol{\theta}(a_i, b_i|\cdot)\) at each step so that it will
not violate the valence constraint.
\(p_\boldsymbol{\theta}(r_i|\cdot)\),
\(p_\boldsymbol{\theta}(\theta_i|\cdot)\),
\(p_\boldsymbol{\theta}(\phi_i|\cdot)\) are mixtures of truncated
Gaussian distributions. The number of mixture is set to be 15, 10 and 5
for \(v^*_1,v^*_2,v^*_3\) respectively. The range of \(r_i\) and
\(\theta_i\) are set to \([0.5, 2.5]\) and \([0, \pi]\). The parameters
of those distribution are calculated using MADE (masked autoencoder for
distribution estimation \cite{Germain.2015}), which is an efficient
architecture for autoregressive models based on masked linear layers.
The architecture for MADE is demonstrated in Figure \ref{fig:made}.

\begin{figure}[!tbh]
\centering
\includegraphics{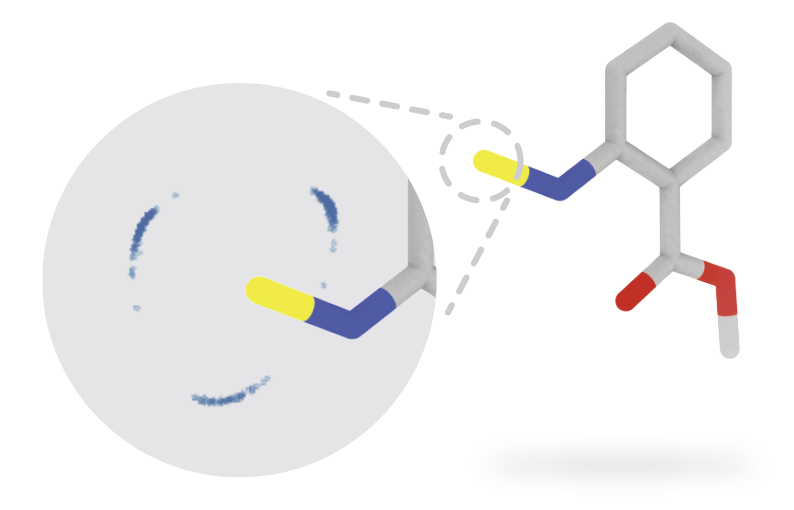}
\caption{The problem of low dimensionality. This figure shows the new
atoms sampled from the model mostly reside inside a 1D ring above the
focused atom.\label{fig:manifold}}
\end{figure}

In practice, the target distribution of atom position usually resides in
a low dimensional submanifold of \(\mathbb{R}^3\) (Figure
\ref{fig:manifold}), and using MADE to directly fit those distributions
will not work well. We adopt the method from a previous work
\cite{Kim.2020} and adds a small error to the target distribution so
that it can be fitted more robustly by MADE. We call this modified model
SoftMADE. Different from the original paper, which samples the noise
level from a uniform distribution: \(c \sim u[a, b]\), we sample the
noise level in two steps:

\[
c_0 \sim u[0, 1]
\]

\[
c = c_\mathrm{max}c_0^\alpha
\]

Where \(c_\mathrm{max}\) is the maximum level of noise, and \(\alpha>1\)
is the parameter controlling the shape distribution. Three conditons are
tested: (1) ordinary MADE; (2) SoftMADE with
\(c_\mathrm{max}=0.2, \alpha=3\); (3) SoftMADE with
\(c_\mathrm{max}=0.4, \alpha=4\). It is find that biger
\(c_\mathrm{max}\) combined with higher \(\alpha\) (the third conditon)
yields better result.

Finally, since the new atoms \(v^*_1,v^*_2,v^*_3\) can be generated in
any order, we use the following corrected likelihood during training:

\[
p(\{v^*_1,v^*_2,v^*_3\}) = 1/3!\sum_{\sigma}{p(v_{\sigma_1}^*, v_{\sigma_2}^*, v_{\sigma_3}^*)}
\]

\hypertarget{decision-making-during-the-connect-operation}{%
\paragraph{Decision making during the ``connect''
operation}\label{decision-making-during-the-connect-operation}}

For each possible action in the ``connect'' operation, we first compute
their unnormalized scores as follows:

\[
\hat{\mathbf{p}}_v^\mathrm{connect}=\mathrm{MLP}_\mathrm{policy}^\mathrm{connect}(\mathbf{h}_{i, v})
\]

\[
\hat{p}_v^\mathrm{skip}=\mathrm{MLP}_\mathrm{policy}^\mathrm{skip}(\mathbf{h}_{i, g})
\]

Where \(\mathrm{MLP}_\mathrm{policy}^\mathrm{connect}\) and
\(\mathrm{MLP}_\mathrm{policy}^\mathrm{skip}\) are fully connected
layers. Those scores are then normalized using softmax:

\[
[\mathbf{p}_v^\mathrm{connect}; p_v^\mathrm{skip}]=\mathrm{softmax}(
[\hat{\mathbf{p}}_v^\mathrm{connect}; \hat{p}_v^\mathrm{skip}]
)
\]

The the values \(\mathbf{p}_v^\mathrm{connect}[b]\) in vector
\(\mathbf{p}_v^\mathrm{connect}\) represents the probability of
connecting the focused atom \(v'\) with \(v\) using a new bond of type
\(b\). The value \(p_v^\mathrm{skip}\) represents the prabability of
skipping the ``connect'' operation and proceeds directly to the
``append'' operation.

\hypertarget{ranking-the-generated-atoms}{%
\paragraph{Ranking the generated
atoms}\label{ranking-the-generated-atoms}}

When ranking the generated atoms, we first calculate an unnormalized
score for each permutation of the new atoms:

\[
\hat{s}_\sigma=\mathrm{MLP}_\mathrm{policy}^\mathrm{rank}([v_{\sigma_1}^*; v_{\sigma_2}^*; v_{\sigma_3}^*])
\]

And then the normalized probability:

\[
p(\sigma) = \frac{\exp{\hat{s}_\sigma}}{\sum_{\sigma'}{\exp{\hat{s}_{\sigma'}}}}
\]

The ranking is then sampled from \(p(\sigma)\).

  \hypertarget{data-collection-and-preprocessing}{%
\subsection{Data collection and
preprocessing}\label{data-collection-and-preprocessing}}

We construct a drug-like subset of ChEMBL \cite{Mendez.2018} for the
training and evaluation of the model. The topological data of all
molecules are downloaded from ChEMBL (version 27) and is then filtered
using a series of criteria:

\begin{itemize}
\tightlist
\item
  Molecules with atom type outside the set \{C, H, O, N, P, S, F, Cl,
  Br, I\} as well as those that do not contain carbon atoms are removed;
\item
  Molecules with the number of heavy atoms outside the range {[}10,
  35{]} are removed;
\item
  Molecules with a QED \cite{Bickerton.2012} value less than 0.5 are
  removed;
\item
  Molecules with ring sizes larger than 7 are removed. Rings in the
  molecule are extracted using RDKit;
\item
  Molecules containing a ring assemble with the number of SSSR(smallest
  set of smallest rings) greater than 4 are removed.
\end{itemize}

After filtering the topological structure, 3D structures are generated
for each molecule using RDKit. The initial 3D embeddings of molecules
are first created using distance geometry and then optimized using the
MMFF94s forcefield. After those processings, we obtain a dataset with 1
million small molecules with 3D structure. The dataset is randomly split
into the training set (4/6), validation set (1/6), and test set (1/6).
The validation set is used during the manual hyperparameter tunning.

\begin{figure}[!tbh]
\centering
\includegraphics[width=\linewidth]{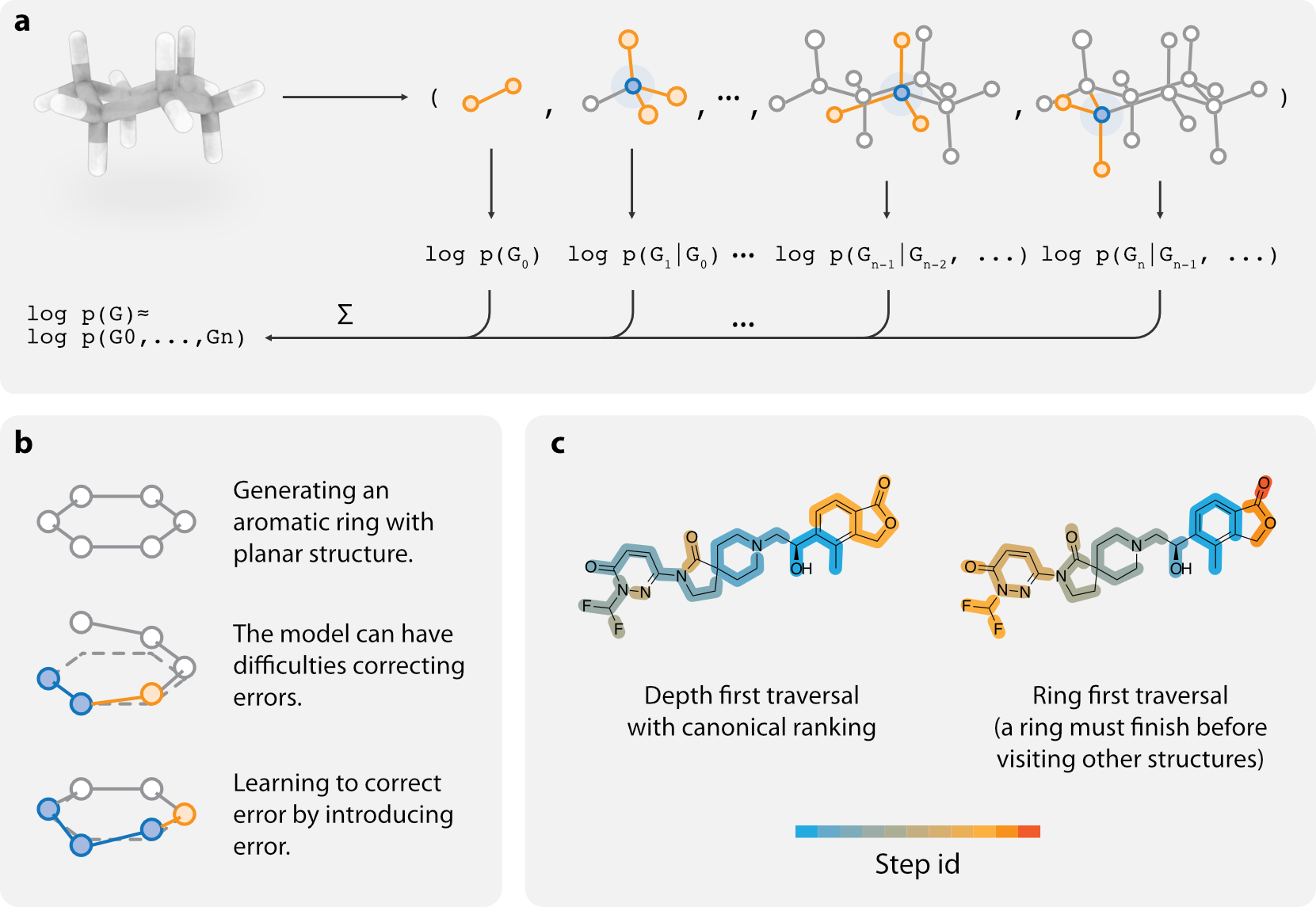}
\caption{Data preprocessing and tricks used to improve model
performance. \textbf{a.} For each data in the training set, we create an
``expert trajectory'' for generating this molecule, and train the model
to imitate this trajectory. \textbf{b.} We find that the model may
suffer from the problem of distribution mismatch, so random errors are
manually added to the training data so that the model can learn to
correct them. \textbf{c.} The image shows the order each atom is
traversed for ordinary (left) and ring first (right) traversal scheme
(blue atoms are traversed first, red atoms are traversed last). We treat
a ring as a basic generation unit and use ring-first traversal to train
the model to close the ring first before generating other
structures.\label{fig:preprocessing}}
\end{figure}

As mentioned in previous sections, the model generates molecules in a
step-by-step manner. To train the model, we need to create an ``expert
trajectory'' for generating each molecule \(G\) in the dataset (Figure
\ref{fig:preprocessing}\textbf{a}):

\[
(G_0, G_1, ..., G_n), \mathrm{where}\ G_1 = \emptyset, G_n = G
\] The model is then trained to imitate this path by maximizing the
log-likelihood:

\[
\log{p_\boldsymbol{\theta}(G)} \approx \sum_{i=1}^{n}\log{p_\boldsymbol{\theta}(G_i|G_0, .., G_i)}
\] We use a method similar to that used previously
\cite{Li.2018,Li.2019} to generate those ``expert trajectories''.
Briefly, the atoms in the molecule are first ranked using a canonical
ranking algorithm in RDKit, and depth-first traversal is performed to
produce a path \((G_0, G_1, ..., G_n)\). Although this method works well
for 2D generative models, it does not produce good results on our 3D
generative model. To improve the performance, we modify the depth-first
traversal algorithm to prioritize the closure of rings (see Figure
\ref{fig:preprocessing}\textbf{c}). We find that this method could
significantly improve the quality of generated samples.

It is also reported that randomized trajectories help the model to
achieve better performance \cite{Li.2018,Arus-Pous.2019}. In this work,
we randomize the trajectory by randomizing the starting position of the
depth-first traversal. We also include data of the model trained using
non-randomized trajectories for comparison.

One of the major issues related to imitation learning is data
distribution mismatch. Specifically, the model only sees correct
``expert trajectories'' during training, and if a mistake happens during
generate, the model may not know how to recover from that error, and
eventually produce invalid results (see Figure
\ref{fig:preprocessing}\textbf{b} for a simple example). Our solution is
to ``simulate'' those errors by adding Gaussian noise to the input by a
certain probability. We use a noise with the standard deviation of 0.1
\(\mathring{A}\) and experimented with the probability of 0.1 and 0.5.
It shows that a 0.5 probability significantly improves the model
performance compared with the 0.1 level.

  \hypertarget{model-training}{%
\subsection{Model training}\label{model-training}}

The model is implemented using PyTorch \cite{Paszke.2019}. Adam
\cite{Kingma.2014} is used to optimize the model parameter. Parameters
\((\beta_1, \beta_2)\) are set to be the default value provided by
PyTorch. The learning rate is initialized to be \(10^{-3}\), and is
decreased by 0.01 for a certain amount of step. Several decay
frequencies are experimented with: every 50 steps, every 100 steps, or
every 200 steps. The batch size for training is set to be 128, and are
trained for a total of 10 epochs. This takes around 3 days to finish.
Training is performed on a single NVIDIA TITAN Xp graphics card.

\hypertarget{hyperparameters}{%
\subsection{Hyperparameters}\label{hyperparameters}}

As can be seen from previous sections, the model proposed here contains
a large number of hyperparameters, including those for model
architecture, data generation, and model training. Considering the long
training time, comprehensive optimization of hyperparameter is
difficult. The hyperparameter selection is further complicated by the
fact that there are a variety of metrics that can be used to evaluate
the model (see section \ref{evaluation}). The best performing model on
one metric does not necessarily performs the best on the other. In this
work, we target 3D MMD as the objective and performs manual
hyperparameter tunning to get to the performance level that is
acceptable for general usage. The result hyperparameter setting is
refered to as the ``standard'' configuration and is what we suggest to
use in future research using this model. We do note that the
optimization process is not comprehensive and we expect that better
performance can be obtained using more dedicated hyperparameter
optimization techniques.

To understand how model performance is affected by a set of
hyperparameter of interest, we perturb those parameters from the
standard configuration to investigate its effect. The set of
hyperparameters that is analyzed are (also summarized in Table
\ref{tab:hyperparameter}):

\begin{itemize}
\tightlist
\item
  The depth and width of DenseNet (see section
  \ref{graph-convolutional-layers});
\item
  The parameters for SoftMADE (see section
  \ref{decision-making-during-the-append-operation});
\item
  The noise added to coordinates of the input structure (see section
  \ref{data-collection-and-preprocessing});
\item
  Whether the ``expert trajectories'' used for training are randomized
  (see section \ref{data-collection-and-preprocessing}).
\end{itemize}

\begin{longtable}[]{@{}lllllll@{}}
\caption{A summary of different hyperparameter configurations whose
performance are reported in this
work.\label{tab:hyperparameter}}\tabularnewline
\toprule
Methods & \shortstack{ Randomized \\trajectory} & c & alpha &
\shortstack{ Input\\noise} & \shortstack{ DenseNet \\architecture} &
\shortstack{ Learning \\rate decay} \\
\midrule
\endfirsthead
\toprule
Methods & \shortstack{ Randomized \\trajectory} & c & alpha &
\shortstack{ Input\\noise} & \shortstack{ DenseNet \\architecture} &
\shortstack{ Learning \\rate decay} \\
\midrule
\endhead
Non-random initialization & No & 0.4 & 4 & 0.5 & Basic & 100 step \\
SoftMADE (low noise) & Yes & 0.2 & 3 & 0.5 & Basic & 100 step \\
No SoftMADE & Yes & 0 & 0 & 0.5 & Basic & 100 step \\
Low input noise & Yes & 0.4 & 4 & 0.1 & Basic & 100 step \\
Shallow DenseNet & Yes & 0.4 & 4 & 0.5 & Shallow & 100 step \\
Narrow DenseNet & Yes & 0.4 & 4 & 0.5 & Narrow & 100 step \\
Slow lr decay & Yes & 0.4 & 4 & 0.5 & Basic & 200 step \\
Fast lr decay & Yes & 0.4 & 4 & 0.5 & Basic & 50 step \\
{\bf Standard configuration} & Yes & 0.4 & 4 & 0.5 & Basic & 100 step \\
\bottomrule
\end{longtable}

  \hypertarget{optimizing-the-speed-of-molecule-generation}{%
\subsection{Optimizing the speed of molecule
generation}\label{optimizing-the-speed-of-molecule-generation}}

\begin{figure}[!tbh]
\centering
\includegraphics{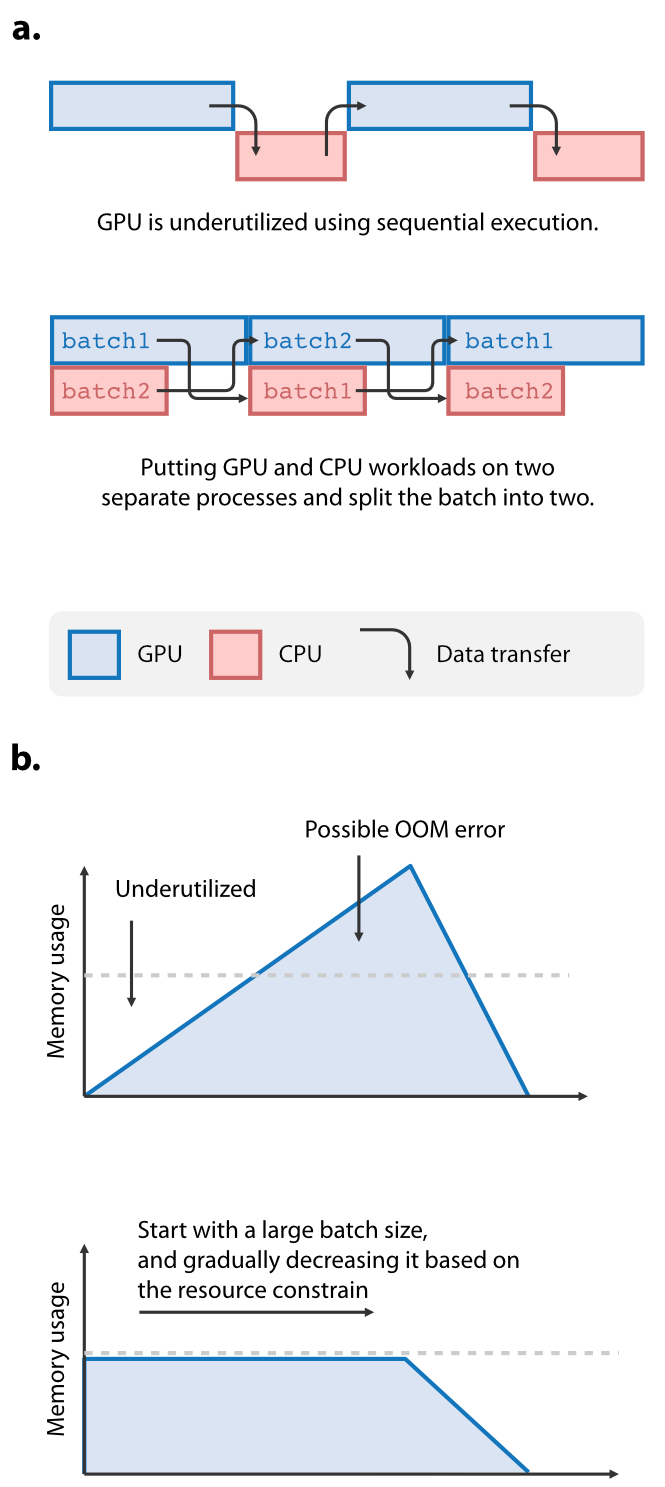}
\caption{Tricks used to accelerate molecule generation: \textbf{a.} CPU
and GPU workloads are separated into two different processes and can be
executed asynchronously by each working on a different batch of data
(batch1 and batch2) at the same time. \textbf{b.} Batch size is adjusted
dynamically to ensure that GPU is fully utilized and does not exceed the
resource constrain. \label{fig:performance}}
\end{figure}

We used several techniques to accelerate the process of molecule
generation. First, many CPU side operations in this model cannot be
implemented efficiently using native python, and we use Numba
\cite{Lam.2015}, a just-in-time compiler for python, to speed up those
codes. The performance benefit is significant, and in our case, the
speedup can be more than 10 times.

Secondly, we use multiprocessing to hide CPU processing latency from
GPU. During molecule generation, we need to move back and forth between
GPU and CPU for action sampling and graph processing. When a large
number of molecules are being generated at the same time, the graph
processing time in the CPU can be significant. Since GPU operation is
blocked by CPU tasks, there will be noticeable performance degradation.
Our solution is to place GPU and CPU in different processes and split
the molecules being generated into two batches. During CPU processing,
GPU can work on the other batch of molecule, thereby increasing the GPU
utilization (Figure \ref{fig:performance}\textbf{a}).

Thirdly, we develop a method to adaptively balancing the number of
molecules generated at each time. During molecule generation, as more
nodes are added to the molecule, the computational cost for the neural
network to process the intermediate structure will be gradually growing.
This will cause a significant performance degration. At the beginning of
generation, the computation burden is low, and GPU is largely
underutilized. When approaching the end of the generation, the
computation cost is significantly higher, and can cause out of memory
(OOM) error (Figure \ref{fig:performance}\textbf{b}). We created a
scheduling method to dynamically adjust the batch size of the generation
task based on the maximum capacity of the GPU. Empirically this has
resulted in significant speed improvement for our model.

After performing those operations, we can achieve a generation speed of
0.008 seconds per molecule in a NVIDIA TITAN Xp card. Note that this
speed is slower than SMILES-based samplers, largely because of the
complexity of the generative workflow and the network architecture. But
we also note that there are still spaces where the performance can be
further improved, like performing quantization or pruning, which will be
explored in the future.

  \hypertarget{evaluation}{%
\subsection{Evaluation}\label{evaluation}}

Several benchmarks have been developed for evaluating 2D generative
models, such as MOSES\cite{Polykovskiy.2020} and
GaucaMol\cite{Brown.2019}, but this type of work is still missing for 3D
models. Here, we assemble a set of evaluation metrics for assessing the
performance of 3D generative models. Emphasis is placed on measuring the
quality of molecule conformations, by investigating various 3D molecular
features. All the following metrics are calculated using 50,000
generated molecules.

\textbf{Output validity and uniqueness}

The two metrics measure the percentage of output molecules that are
chemically valid and structurally unique. The validity is measured by
calculating the percentage of generated molecules that pass the RDKit
sanitization check (\%valid). Although this metric measures the quality
of topological structure, we find that in practice it is also a good
proxy for the quality of 3D structures. This is because that many
serious 3D errors occurring during the generation will eventually lead
to invalid topological structures. The uniqueness is measured by
calculating the percentage of unique structures among outputs (\%uniq).
This can be used to detect whether the model has been overtrained or has
collapsed to a single mode.

\textbf{Distribution of molecular properties}

Investigating the distribution of molecular properties is a good way to
intuitively assess the quality of generated samples. The mean and
standard deviation of each property is reported, as well as the
visualization using kernel density estimation. The properties
investigated include regular topological features (molecular weight,
LogP, the number of hydrogen donors and acceptors, the number of
rotatable bonds, QED) as well the following 3D features:

\begin{itemize}
\tightlist
\item
  Normalized PMI ratios (NPRs) \cite{Sauer.2003}: This is a shape
  descriptor composed of two components
  \((NPR1, NPR2)=(I_1/I_3, I_2/I_3)\), where \(I_1, I_2, I_3\) are
  principal moments of inertia sorted by ascending magnitude. The point
  \((1, 1)\), \((0.5, 0.5)\), \((0, 1)\) corresponds to the archetypes
  of sphere, disk and rod, giving this descriptor high interpretability.
  NPRs are calculated using the implementation in RDKit.
\item
  Solvent accessible surface areas (SASA): SASA is an important
  molecular descriptor measuring the contact area between the molecule
  and the solvent. We report the distribution of polar and total SASA
  calculated using the package FreeSASA\cite{Mitternacht.2016}.
\end{itemize}

\textbf{Maximum mean discrepancy}

We use maximum mean discrepancy (MMD) to give a quantitative measurement
of the difference between the distribution of generated and real
samples. Given a kernel function \(\kappa(\cdot, \cdot)\), the MMD
between two distributions can be estimated as :

\[
MMD = \frac{1}{N(N - 1)}\sum_{i=1}^N{\sum_{j=1, j \ne i}^N{\kappa(x_i, x_j)}}
+
\frac{1}{M(M - 1)}\sum_{i=1}^M{\sum_{j=1, j \ne i}^M{\kappa(y_i, y_j)}}
-
\frac{2}{MN}\sum_{i=1}^N\sum_{j=1}^M\kappa(x_i, y_j)
\]

Where \(\{x_i\}_{i=1}^N\) are sampled from the real distribution and
\(\{y_i\}_{i=1}^M\) sampled from the generative model. This method has
been previously used to assess the performance of 2D molecular
generative models \cite{Li.2019}, and can be easily applied to 3D models
by changing \(\kappa\) to measure 3D molecular similarity.

In this work, we use two types of kernel function to calculate MMD: (1)
the Tanimoto similarity of 2D Morgan fingerprint (1024 bit, radius of 2)
and (2) the Manhattan distance of USRCAT
fingerprint\cite{Schreyer.2012}. The two MMD measures the topological
and 3D discrepancy between generated and real samples. Note that for the
3D kernel function, a more accurate choice might be using the
shape-based alignment method\cite{Grant.1996}. But since the
computational complexity of MMD calculation is \(O(\max(M, N)^2)\), and
shape alignment involves optimization for each molecule pair, this will
not be feasible for our task. USRCAT is an alignment-free method with
precalculated fingerprint and is more suitable for MMD calculation.

The calculation of MMD is parallelized in GPU using Cupy \cite{8qv}. For
topological fingerprint, we store the 1024 bit fingerprint into 32
uint32 integers and utilize bitwise operations to speed up the
calculation of Tanimoto coefficient.

\textbf{Precision and recall}

Although MMD can reliably quantify the discrepancy between
distributions, its value is difficult to interpret. Ideally, we want the
information about:

\begin{itemize}
\item
  What percentage of generated samples are realistic;
\item
  What percentage of real data distribution can be covered by the
  generative model;
\end{itemize}

We call the two metrics precision and recall for the generative model.
Precision can be used to measure the sample quality, and recall can be
used to assess mode coverage. Since they are both percentage values,
they are more interpretable than MMD.

The definition of the two metrics follows the previous work
\cite{Kynkaanniemi.2019}. First of all, we define the space covered by a
probability distribution from its samples \(X=\{x_i\}_{i=1}^N\) as:

\[
\Phi(X)=\{x|\exists\ x_i \in X\ \mathrm{s.t.}\ d(x-x_i)\le d(N_k(x_i, X)-x_i)\}
\]

Where \(N_k(x_i, X)\) denotes the \(k\)-th nearest neighbor of \(x_i\)
in the dataset \(X\), and \(d(\cdot,\cdot)\) measures the distance
between two data point. Given the real data \(X=\{x_i\}_{i=1}^N\) and
generated data \(Y=\{y_i\}_{i=1}^M\), precision and recall are defined
as:

\[
P=\frac{|Y \cap \Phi(X)|}{|Y|}
\] \[
R=\frac{|X \cap \Phi(Y)|}{|X|}
\]

Similar to MMD, a 2D and 3D version of precision and recall can be
calculated using Morgan fingerprint and USRCAT. The value of \(k\) is
set to be 3.

\textbf{Validity of local geometries}

We check the correctness of local 3D structures in generated molecules
by examining the distributions of bond lengths and bond angles. More
specifically, we group the bond lengths and angles by its environment
key, calculate the mean and standard deviation within each group, and
compare them between generated and test set molecules. The environment
key for the bond length contains the bond type and the type of its two
adjacent atoms: \((a_u, a_v, b_{uv})\), while that for a bond angle
contains the type and hybridization state of the central atom:
\((h_u, a_u)\). Groups containing less than 1000 data points are removed
from the evaluation.

We also check the distribution of torsion angles in generated and test
set molecules. Quads of atoms a-b-c-d are matched using the SMARTS
pattern provided in the previous work \cite{Riniker.2015} and torsion
angles are calculated from the matched coordinates. Patterns that give
less than 1000 matches are discarded. MMD values of the torsion
distribution in each pattern are computed.

The about evaluation aims at giving a qualitative look about the quality
of local geomerty, and the comparison is only performed on the model
trained with standard configuration. We use RMSD (Root Mean Standard
Deviation) for a more quantitative evaluation, as described below.

\textbf{RMSD}

For each generated molecule, we optimize the conformation using MMFF94s
force field and calculate the RMSD of heavy atoms between the original
and optimized structure.

To give a context on how the model performs in this metric, we perform
the same calculation on the ETKDG method \cite{Riniker.2015}, which is a
conformation generation method for small molecules that aims to provide
a faster alternative for forcefield-based minimization. For each
molecule in the test set, we generate an initial conformation using
ETKDG, optimize it using MMFF94s, and calculate the RMSD for
conformation before and after optimization. The average RMSD is then
compared with that of the generative model.

Note however that since ETKDG and deep generative model is developed to
solve two completely different problems, the comparison of RMSD can not
tell which method is better or worse. Nonetheless, this comparison
should give us an idea about the overall level of quality of the
generated conformations.

\hypertarget{combining-l-net-with-mcts-for-structure-based-molecule-design}{%
\subsection{Combining L-Net with MCTS for structure-based molecule
design}\label{combining-l-net-with-mcts-for-structure-based-molecule-design}}

The model proposed in this work can be conveniently combined with other
techniques such as reinforcement learning to achieve molecular design
based on a given objective. As a proof of concept, we combine L-Net with
Monte Carlo tree search (MCTS) and test its ability in the problem of
structure-based molecule design. Previous works have combined MCTS with
2D generative models in the object-directed design of
molecules\cite{Yang.2017d9g}, but to our knowledge, the combination of
MCTS with 3D generative models has not yet been reported.

Technically, we use L-Net as the rollout policy for MCTS, and Maximum
Entropy for Tree Search (MENTS)\cite{xiao2019maximum} for tree policy.
We periodically perform an ``exploit'' step that samples directly from
the distribution defined by the Q-values. For each sample, local
optimization is performed using smina\cite{Koes.2013}, and the resulted
(minus) affinity score is treated as the reward function. Note that the
optimization is only performed locally not globally, so the evaluation
process can be done extremely fast. Water molecules are striped from the
structure before scoring using smina. To better utilize the
computational power of GPU, leaf-level and tree-level parallelization is
introduced to MCTS, similar to that done by Chaslot et
al.\cite{Chaslot.2008}.

\begin{figure}[!tbh]
\centering
\includegraphics[width=\linewidth]{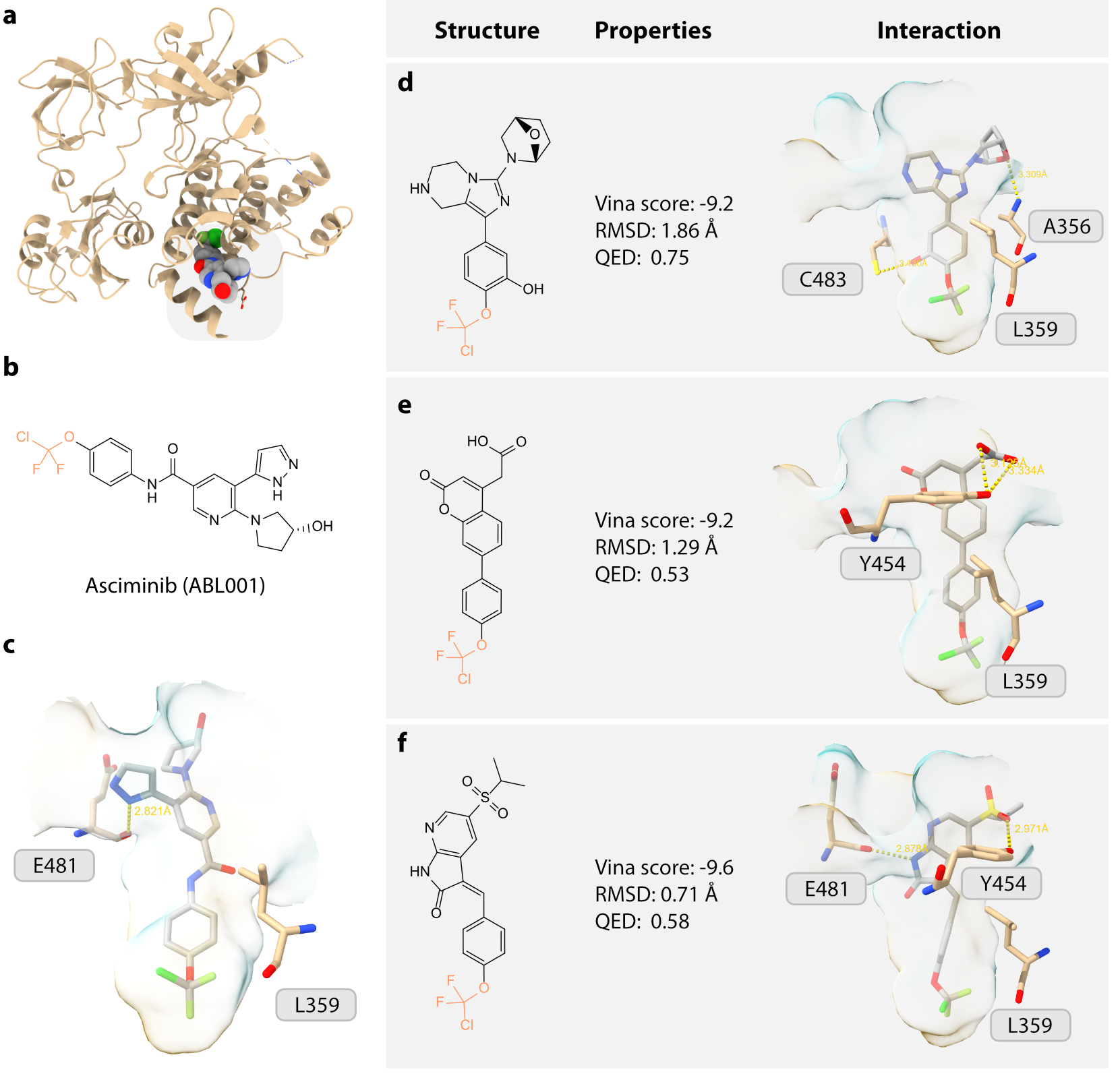}
\caption{\textbf{a.} The structure of asciminib binding to the
allosteric pocket in ABL1. \textbf{b.} The topological structure of
asciminib, with the seed structure highlighted. \textbf{c.} A closer
look of asciminib inside the binding pocket. \textbf{d-f.} The
structures of three selected molecules from generated samples and their
binding poses predicted by smina.\label{fig:5mo4}}
\end{figure}

The case target that we use to test the model is the Tyrosine-protein
kinase ABL1, which is related to chronic myelogenous leukemia (CML).
Instead of the frequently targeted ATP-site, here we focus on the
allosteric myristate binding site due to its potential advantages in
selectivity. However, using binding affinity scores from smina to guide
the design of allosteric inhibitors might be problematic. Indeed,
previous works have reported several binders of the ABL1 allosteric site
with no inhibition activity \cite{Schoepfer.2018}. Instead of trying to
generate the entire structure using L-Net, we use the \(CClF_2O-\) group
in asciminib (a known active molecule currently under clinical trial,
see Figure \ref{fig:5mo4}\textbf{b}) as seed structure. Previous reports
have shown that this functional group is essential for the inhibition
activity of asciminib\cite{Schoepfer.2018}. Based on this seed
structure, we then use the L-Net with MCTS to generate and optimize the
rest of the structure for higher binding affinity.

  \hypertarget{results-and-discussion}{%
\section{Results and Discussion}\label{results-and-discussion}}

\hypertarget{generated-samples-validity-and-uniqueness}{%
\subsection{Generated samples, validity and
uniqueness}\label{generated-samples-validity-and-uniqueness}}

\begin{figure}[!tbh]
\centering
\includegraphics[width=\linewidth]{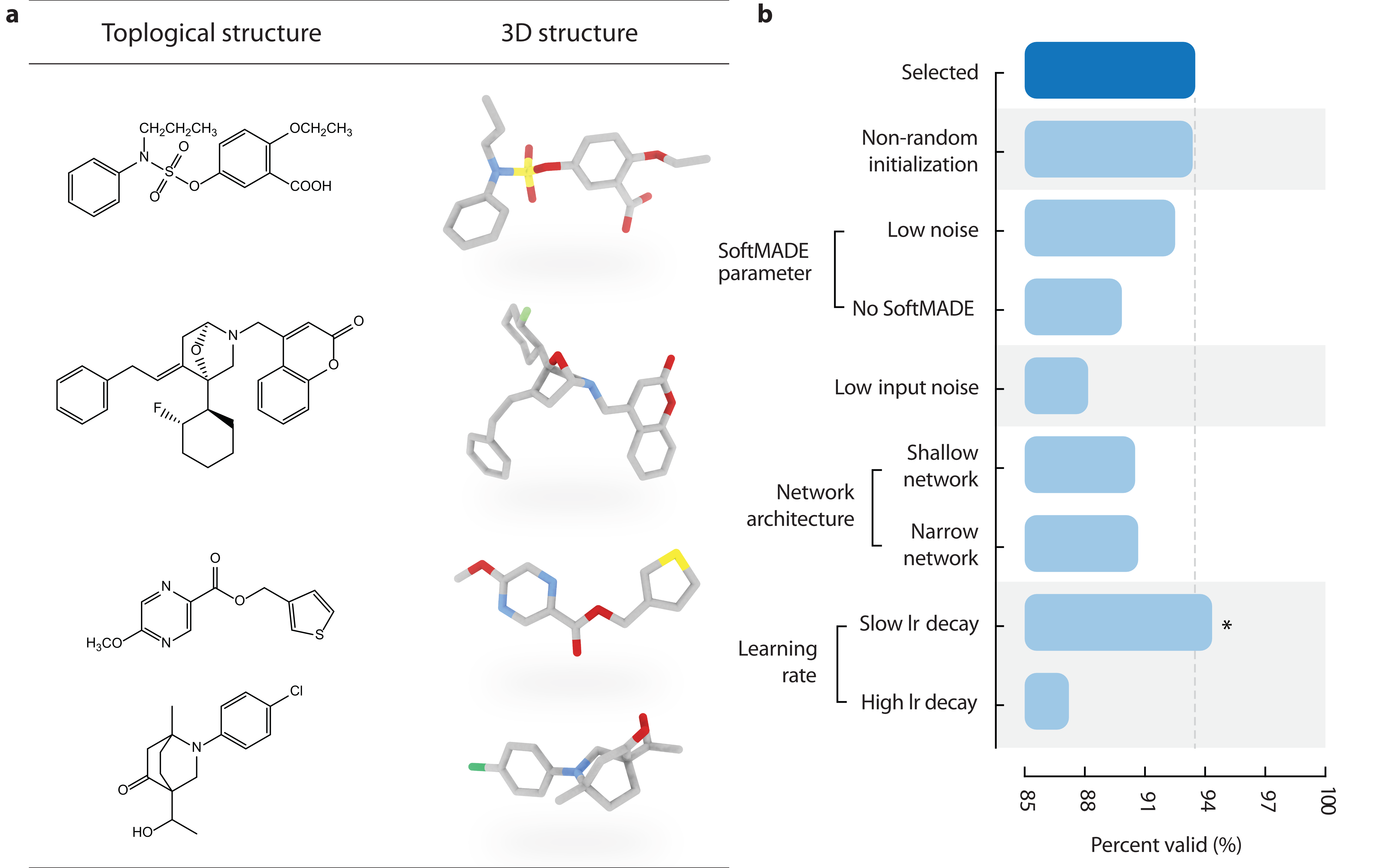}
\caption{\textbf{a.} Several randomly generated samples from LNet;
\textbf{b.} The percentage of output molecules with valid chemical
structures, depending on the different selection of hyperparameters. The
star indicates the best performing hyperparameter
selection.\label{fig:valid}}
\end{figure}

Several randomly generated samples are shown in Figure
\ref{fig:valid}\textbf{a}, with topological and 3D structures (rendered
using ChimeraX \cite{Goddard.2018}). Visual inspection shows that those
molecules have correct local geometry. For example, sp3 and sp2
hybridized atoms correctly adopt tetrahedral and planar geometry, and
aromatic systems correctly form planar structures.

\begin{longtable}[]{@{}lll@{}}
\caption{The performance of LNet, measured in terms of \%valid and
\%uniq, with different hyperparameters. The star indicates the best
performing hyperparameter selection.\label{tab:valid}}\tabularnewline
\toprule
Method & \%Valid & \%Uniq \\
\midrule
\endfirsthead
\toprule
Method & \%Valid & \%Uniq \\
\midrule
\endhead
Non-random initialization & 93.4\% & 98.7\% \\
SoftMADE (low noise) & 92.5\% & 99.0\% \\
No SoftMADE & 89.8\% & 89.7\% \\
Low input noise & 88.2\% & 97.4\% \\
Shallow DenseNet & 90.5\% & 98.9\% \\
Narrow DenseNet & 90.7\% & 98.6\% \\
Slow decay & 94.3\% (*) & 98.2\% \\
Fast decay & 87.2\% & 99.0\% \\
{\bf Standard configuration} & 93.5\% & 99.1\% (*) \\
\bottomrule
\end{longtable}

Table \ref{tab:valid} demonstrates the ability of LNet to generate valid
(measured by \%valid) and unique (measured by \%uniq) molecules. Our
model can achieve as high as 94.3\% of output validity. For \%uniq, the
reported values are generally approaching 100\%, indicating no
overtraining or mode collapse. Note in the result that using SoftMADE
(row 9 v.s. row 2 and row 3) significantly improve the model's
performance in terms of \%valid, indicating that naive MADE indeed
suffers from the problem of low dimensional manifold (discussed in
Section \ref{decision-making-during-the-append-operation}) in our task.
The addition of input error (row 9 v.s. 4) also helps to improve the
performance of \%valid, likely due to the reason discussed in Section
\ref{data-collection-and-preprocessing}. Note however that randomized
trajectory (row 9 v.s. row 1) does not significantly improve model
performance. This might because we only randomized the selection of the
first atom. We find that using techniques such as temperature sampling
can further improve the \%valid value. However, we find that this method
would significantly reduce other performance metrics such as MMD, and is
therefore not adopted.

Other works in the 3D generative model have reported much lower \%valid
values. Gebauer et al.~have reported a 77.07\% result on \%valid
\cite{Gebauer.2019} using QM9, which is structurally much simpler than
the dataset we use. Ragoza et al.~reported a closer 90\% result on
\%valid \cite{Ragoza.2020}, using molecules from the MolPort dataset,
but their method is not end-to-end and requires separate atom placement
and bond order assignment algorithms. We also note that the \%valid
result for our model is lower than most 2D generative models, even
simple SMILES-based methods. This phenomenon is also reported in the
result of G-SchNet \cite{Gebauer.2019}. The major reason is that for our
model, not only topological errors will contribute to invalid molecules,
but also 3D ones. In fact, most 3D errors will eventually be converted
to topological errors in our model. This means that 3D generative tasks
for molecule exhibits higher difficulty compared with 2D tasks.

  \hypertarget{distribution-of-molecular-properties}{%
\subsection{Distribution of molecular
properties}\label{distribution-of-molecular-properties}}

{\tiny
\begin{longtable}[]{@{}lllllllllllll@{}}
\caption{Distribution of 2D molecular properties among generated
molecules using different
hyperparameters.\label{tab:prop2d}}\tabularnewline
\toprule
& MW & & LogP & & HBA & & HBD & & ROTB & & QED & \\
\midrule
\endfirsthead
\toprule
& MW & & LogP & & HBA & & HBD & & ROTB & & QED & \\
\midrule
\endhead
Methods & mean & std. & mean & std. & mean & std. & mean & std. & mean &
std. & mean & std. \\
\hline
Non-random initialization & 328.5 & 86.8 & 2.89 & 1.45
& 4.00 & 1.67 & 1.25 & 0.95 & 4.26 & 2.25 & 0.683 & 0.148 \\
SoftMADE (low noise) & 332.0 & 87.6 & 2.87 & 1.47 &
4.06 & 1.71 & 1.29 & 0.99 & 4.18 & 2.18 & 0.675 & 0.152 \\
No SoftMADE & 329.0 & 94.3 & 2.86 & 1.55 & 4.07 & 1.80
& 1.20 & 0.98 & 4.20 & 2.28 & 0.664 & 0.156 \\
Low input noise & 356.4 & 119.1 & 2.66 & 1.68 & 4.72 &
2.11 & 1.47 & 1.10 & 4.76 & 2.79 & 0.588 & 0.182 \\
Shallow DenseNet & 338.9 & 95.3 & 2.84 & 1.55 & 4.21 &
1.78 & 1.31 & 1.01 & 4.29 & 2.30 & 0.662 & 0.158 \\
Narrow DenseNet & 329.0 & 93.3 & 2.74 & 1.52 & 4.17 &
1.79 & 1.20 & 0.98 & 4.22 & 2.26 & 0.669 & 0.154 \\
Slow decay & 332.8 & 90.4 & 2.78 & 1.48 & 4.21 & 1.75 &
1.35 & 1.02 & 4.51 & 2.35 & 0.664 & 0.152 \\
Fast decay & 339.5 & 94.3 & 2.88 & 1.57 & 4.17 & 1.78 &
1.26 & 1.01 & 4.54 & 2.40 & 0.655 & 0.162 \\
{\bf Standard configuration} & 338.0 & 91.7 & 2.91 & 1.55 &
4.12 & 1.75 & 1.26 & 0.99 & 4.40 & 2.28 & 0.665 & 0.156 \\
\hline
Validation & 345.0 & 69.7 & 3.05 & 1.30 & 4.13 & 1.55 & 1.19 & 0.91 &
4.30 & 2.04 & 0.700 & 0.117 \\
Test & 344.8 & 69.7 & 3.05 & 1.29 & 4.13 & 1.56 & 1.20 & 0.90 & 4.30 &
2.04 & 0.701 & 0.116 \\
\bottomrule
\end{longtable}
}

\begin{figure}[!tbh]
\centering
\includegraphics[width=\linewidth]{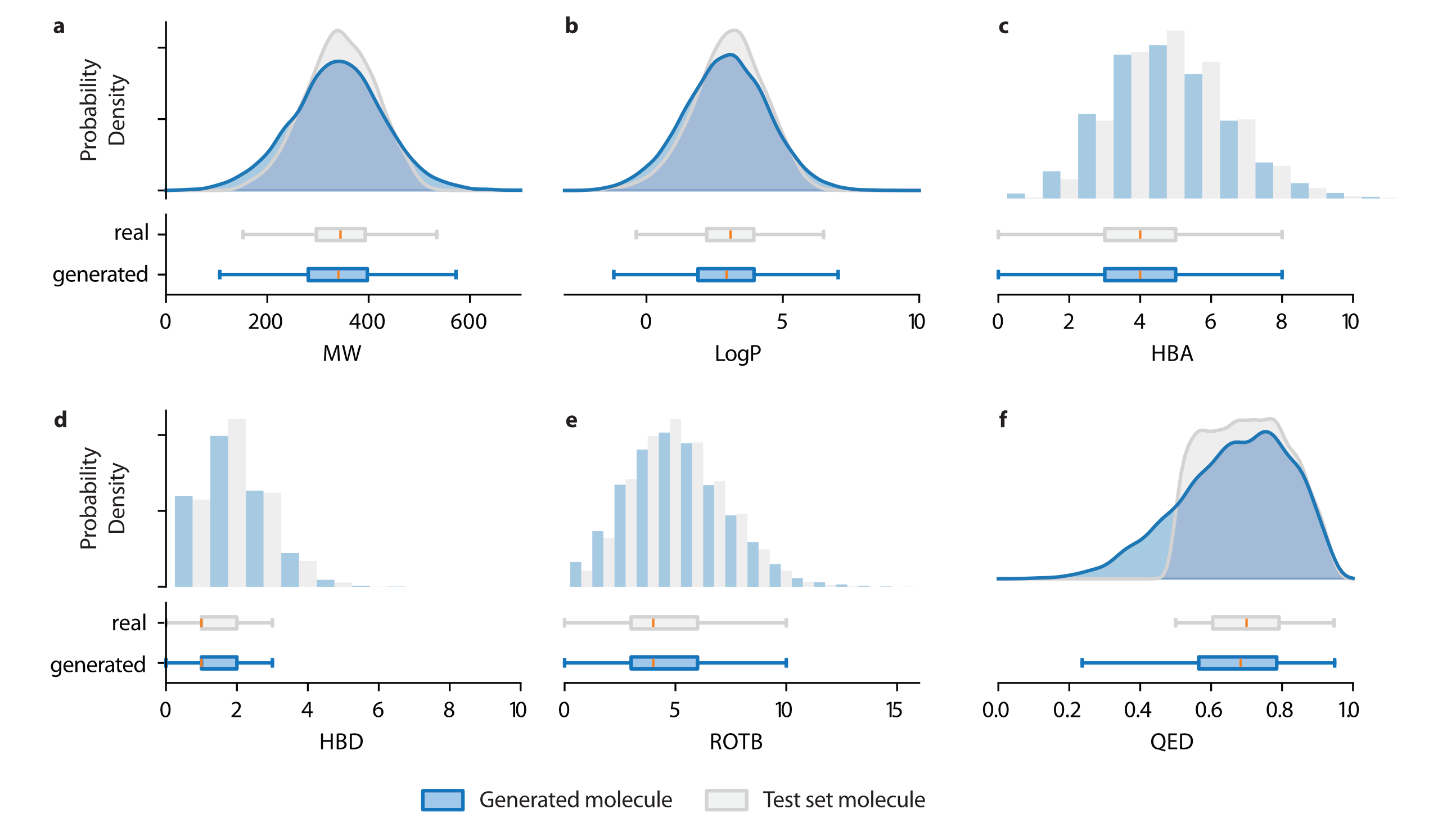}
\caption{The distribution of 2D molecular properties of generated
molecules and test set molecules. \textbf{a.} Molecular weight (MW).
\textbf{b.} LogP. \textbf{c.} The number of hydrogen bond acceptors
(HBA). \textbf{d.} The number of hydrogen bond donors (HBD). \textbf{e.}
The number of rotatable bonds (ROTB). \textbf{f.} Druglikeness (QED).
Generated molecules are shown in blue, and test set molecules are shown
in grey.\label{fig:prop2d}}
\end{figure}

We first investigate the topological properties of generated samples.
Table \ref{tab:prop2d} shows the mean and standard deviation of various
topological properties among the generated samples and samples in the
validation and test set, using different combinations of
hyperparameters. Figure \ref{fig:prop2d} give a visualized presentation
of the distribution for generated (blue) and test set molecules (grey),
either using the probability value from kernel density estimation or
using histogram from continuous properties. Horizontal box plots are
also given (below the histograms).

The first thing that we may notice is that the property of generated
molecule tends to be more spread than that in the test set. This can be
demonstrated by the standard deviation of each property shown in Table
\ref{tab:prop2d}. This may be an indication that the model is
prioritizing mode coverage over precision, which is also supported by
the result of precision and recall values (see Section
\ref{precision-and-recall}). The mean values of each property match
quite well, for molecule weight, the difference is less than 10(row 9
v.s. row 10 and 11). It is also noticed that a large discrepancy is
found between the generated and real QED distributions (Figure
\ref{fig:prop2d} \textbf{f}). This is the result of a hard cutoff value
of 0.5 during data selection.

\begin{longtable}[]{@{}lllllllll@{}}
\caption{Distribution of 3D molecular properties among generated
molecules using different
hyperparameters.\label{tab:prop3d}}\tabularnewline
\toprule
& Total SASA & & Polar SASA & & NPR1 & & NPR2 & \\
\midrule
\endfirsthead
\toprule
& Total SASA & & Polar SASA & & NPR1 & & NPR2 & \\
\midrule
\endhead
Methods & mean & std. & mean & std. & mean & std. & mean & std. \\
\hline
\shortstack[l]{Non-random\\initialization} & 514.9 & 100.1 & 126.2 &
57.6 & 0.252 & 0.130 & 0.859 & 0.092 \\
\shortstack[l]{SoftMADE\\(low noise)} & 518.6 & 100.8 & 129.4 & 58.3 &
0.252 & 0.130 & 0.859 & 0.093 \\
\shortstack[l]{No\\SoftMADE} & 514.6 & 107.5 & 128.1 & 60.3 & 0.261 &
0.131 & 0.854 & 0.094 \\
\shortstack[l]{Low input\\noise} & 540.3 & 135.9 & 148.8 & 66.9 & 0.278
& 0.134 & 0.848 & 0.095 \\
\shortstack[l]{Shallow\\DenseNet} & 524.7 & 107.6 & 133.3 & 59.5 & 0.263
& 0.134 & 0.856 & 0.093 \\
\shortstack[l]{Narrow\\DenseNet} & 514.3 & 106.5 & 128.8 & 59.7 & 0.265
& 0.132 & 0.854 & 0.094 \\
\shortstack[l]{Slow\\decay} & 522.2 & 104.5 & 132.1 & 58.7 & 0.256 &
0.132 & 0.858 & 0.093 \\
\shortstack[l]{Fast\\decay} & 525.3 & 106.9 & 129.2 & 59.6 & 0.271 &
0.136 & 0.856 & 0.093 \\
\shortstack[l]{\bf Standard\\ \bf configuration} & 565.9 & 107.9 & 129.0 & 60.2 &
0.263 & 0.133 & 0.857 & 0.093 \\
\hline
Validation & 579.5 & 88.5 & 129.5 & 55.9 & 0.232 & 0.125 & 0.869 &
0.093 \\
Test & 537.8 & 84.0 & 128.6 & 53.5 & 0.232 & 0.124 & 0.869 & 0.093 \\
\bottomrule
\end{longtable}

\begin{figure}[!tbh]
\centering
\includegraphics[width=\linewidth]{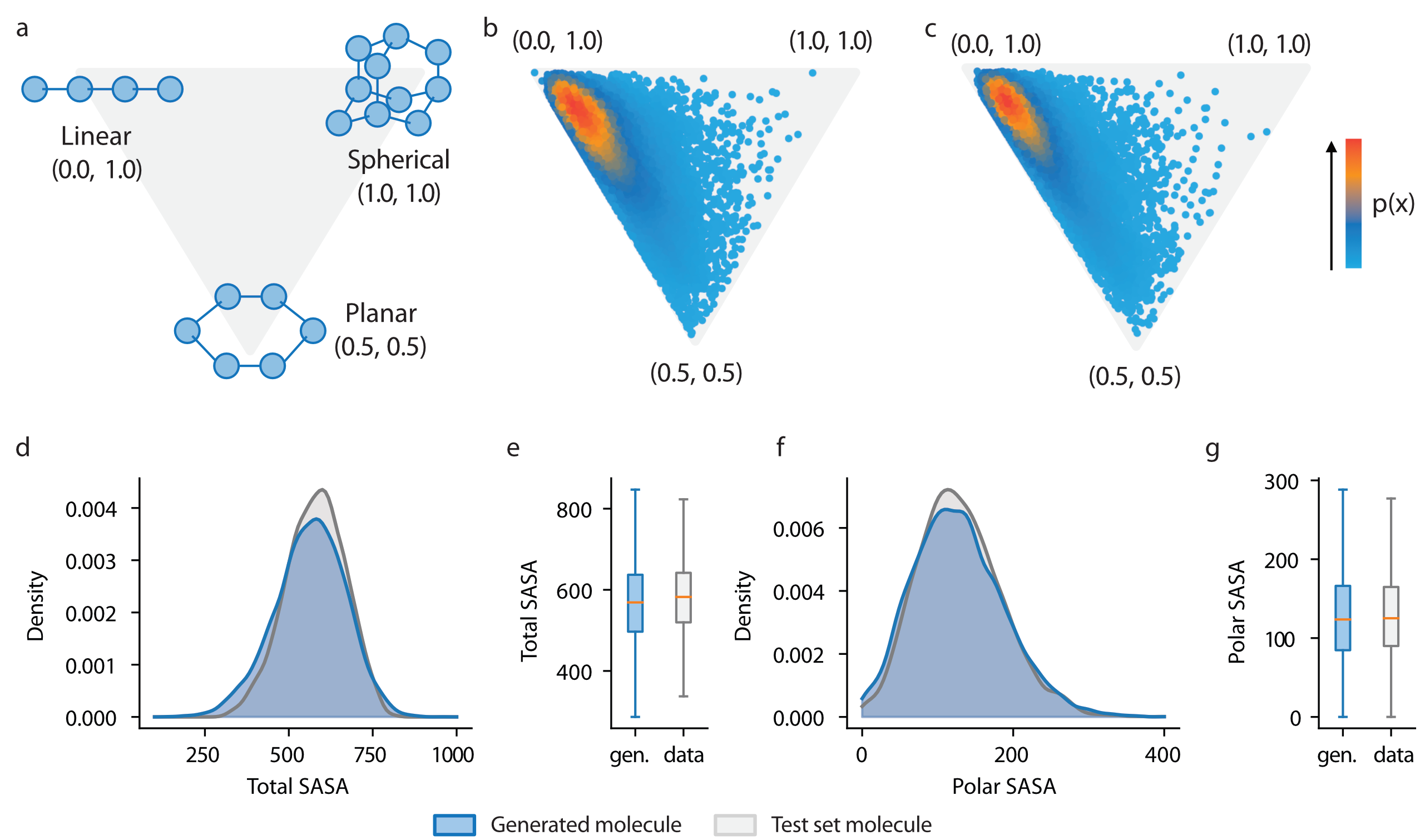}
\caption{The distribution of 3D molecular properties. \textbf{a.}
Interpreting the NPR shape descriptors. \textbf{b-c.} Comparing the
distribution of NRPs between generated (\textbf{b}) and test set
(\textbf{c}) molecules. \textbf{d-g.} Comparison of Total SASA and Polar
SASA between generated and test set molecules (\textbf{d,f}: Kernel
density estimation; \textbf{e, g}: Box plot; Blue: generated molecules;
Grey: test set molecules).\label{fig:prop3d}}
\end{figure}

Table \ref{tab:prop3d} and Figure \ref{fig:prop3d} shows the
distribution of 3D properties among generated, validation set and test
set molecules. Total and polar SASA measures the surface accessible
surface area of molecules, and NPR1 and NPR2 measures the overall shape
of the molecule (linear, planar, or spherical, see Figure
\ref{fig:prop3d}\textbf{a}). Similar to that discovered in the
distribution of 2D properties, generated samples generally exhibit a
wider probability distribution of 3D properties. This discrepancy is no
so evident when inspecting the visualized data in Figure
\ref{fig:prop3d}. For both generated and test set molecules, the shape
distribution tends to gather around the ``linear'' corner of the
triangle, while slightly tilled towards the ``planar'' corner.

  \hypertarget{mmd-and-diversity}{%
\subsection{MMD and diversity}\label{mmd-and-diversity}}

\begin{figure}[!tbh]
\centering
\includegraphics[width=\linewidth]{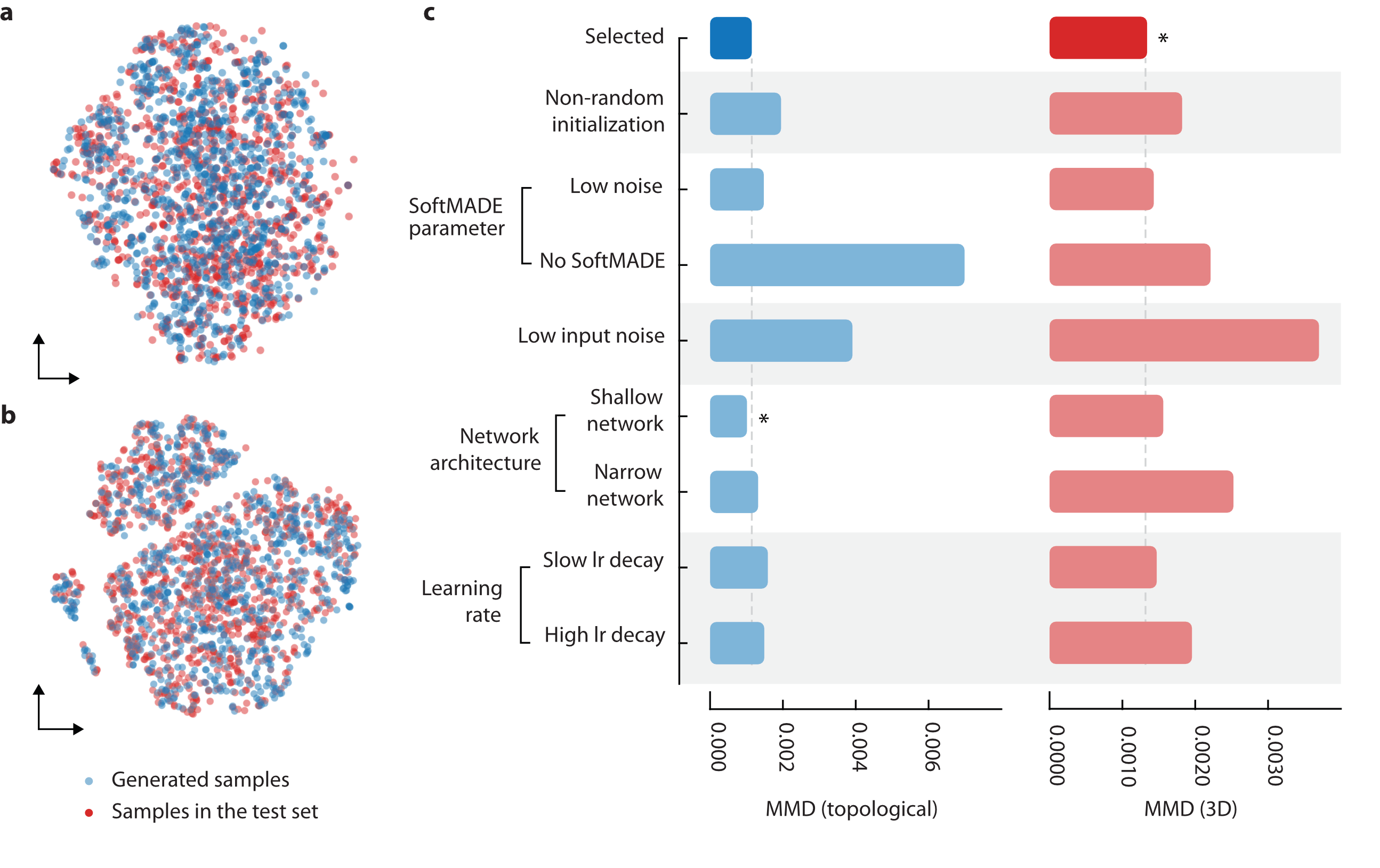}
\caption{\textbf{a-b.} A t-SNE visualization of the distribution of
Morgan (\textbf{a}) and USRCAT (\textbf{b}) fingerprint in two dimension
space (Blue: generated samples; Red: samples in the test set).
\textbf{c.} The Topological and 3D MMD result in the form of bar plot. A
star indicates the best performing combination of perameters in that
metric.\label{fig:mmd}}
\end{figure}

After a qualitative inspection of sample quality using \%valid and
property distribution, we move to a more quantitative approach at
measuring the discrepancy between the distribution of generated and real
molecules. Topological and 3D MMD values are calculated from Morgan
fingerprints and USRCAT respectively. A visualization of the
distribution of those fingerprint is demonstrated in Figure
\ref{fig:mmd} \textbf{a, b}. The high dimensional representation of
molecule in fingerprint is embedded into two-dimensional space using
TSNE \cite{van2008visualizing} (implemented in scikit-learn
\cite{Pedregosa.2012}). It is found that generated samples and samples
in the test set are evenly mixed in the two-dimensional space, therefore
indicating that there is not significant mismatch between the generated
and test set samples.

The result of Topological and 3D MMD values are demonstrated in Figure
\ref{fig:mmd} \textbf{c} and Table \ref{tab:mmd}. We start by inspecting
the result of topological MMD. It is quickly discovered that SoftMADE
and input noise significantly improve the result of topological MMD.
This may be contributed by the fact that SoftMADE and input noise reduce
errors during the generation. Errors may happen more frequently on
topologically complexed molecules, thereby causing the discrepancy
between 2D structure distribution. Other hyperparameters do not have a
significant effect on topological MMD. Interestingly, having shallow
neural network architecture even slightly improves the MMD value.
However, using a shallow network does have a significantly negative
impact on other metrics (such as \%valid), so it should not be the
architecture of choice for this task. For 3D MMD, the trend is similar
for different hyperparameter selections.

{\tiny
\begin{longtable}[]{@{}lllllllll@{}}
\caption{The result of topological and 3D MMD, sample diversity,
precision and recall for each combination of hyperparameters.
\label{tab:mmd}}\tabularnewline
\toprule
& 3D & & & & Topological & & & \\
\midrule
\endfirsthead
\toprule
& 3D & & & & Topological & & & \\
\midrule
\endhead
Methods & MMD & Diversity & Precision & Recall & MMD & Diversity &
Precision & Recall \\
\hline
Non-random initialization & 0.00182 & 0.154 & 83.6\%
(*) & 87.8\% & 0.00195 & 0.160 (*) & 51.4\% (*) & 71.8\% \\
SoftMADE (low noise) & 0.00143 & 0.155 (*) & 83.1\% &
88.2\% & 0.00148 & 0.158 & 46.6\% & 74.8\% \\
No SoftMADE & 0.00221 & 0.131 & 81.8\% & 88.3\% &
0.00699 & 0.158 & 39.2\% & 76.8\% \\
Low input noise & 0.00370 & 0.143 & 76.0\% & 88.6\% &
0.00391 & 0.148 & 28.8\% & 84.1\% (*) \\
Shallow DenseNet & 0.00156 & 0.151 & 81.5\% & 88.2\% &
0.00101 (*) & 0.156 & 41.8\% & 76.8\% \\
Narrow DenseNet & 0.00253 & 0.148 & 81.7\% & 88.4\% &
0.00132 & 0.157 & 41.6\% & 76.6\% \\
Slow decay & 0.00147 & 0.151 & 81.9\% & 88.3\% &
0.00159 & 0.155 & 44.9\% & 76.5\% \\
Fast decay & 0.00196 & 0.150 & 80.9\% & 88.6\% (*) &
0.00149 & 0.156 & 38.7\% & 79.0\% \\
{\bf Standard configuration} & 0.00134 (*) & 0.152 & 81.9\% &
88.3\% & 0.00115 & 0.157 & 43.1\% & 78.0\% \\
\hline
Validation & - & 0.157 & - & - & - & 0.157 & - & - \\
Test & - & 0.157 & - & - & - & 0.157 & - & - \\
\bottomrule
\end{longtable}
}

  \hypertarget{precision-and-recall}{%
\subsection{Precision and recall}\label{precision-and-recall}}

\begin{figure}[!tbh]
\centering
\includegraphics[width=\linewidth]{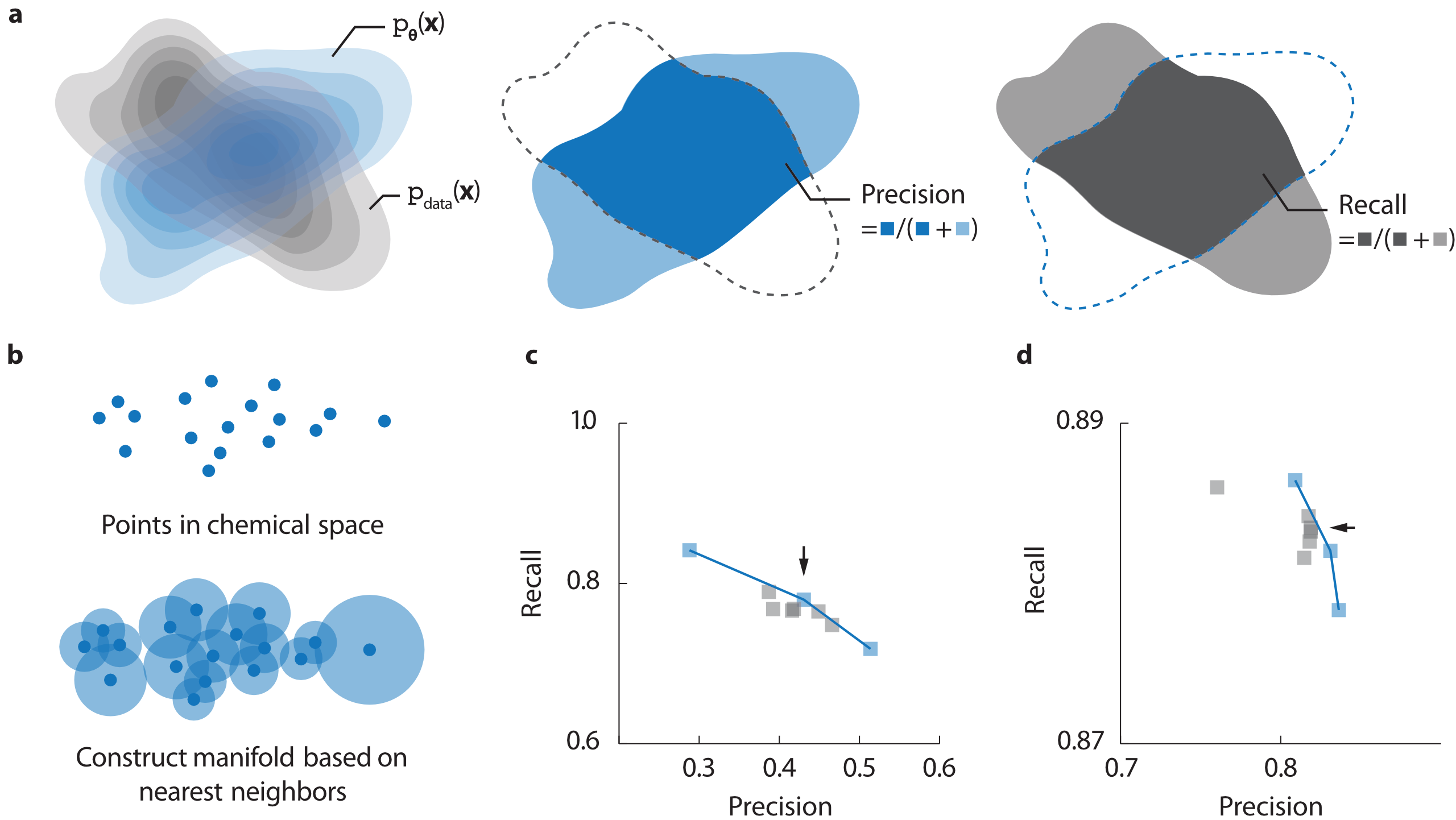}
\caption{\textbf{a.} The method of calculating precision and recall for
generative models. Precision is calculated as the percentage of
generated samples covered by real samples, while recall is calculated as
the precentage of real samples covered by generated samples. \textbf{b.}
The space covered by a probability distribution is estimated from its
samples, using k-nearest neighbors; \textbf{c-d.} The precision and
recall values calculated using topological (\textbf{c}) and 3D
fingerprint (\textbf{d}), the blue line indicates the Pareto frontier,
the black arrow indicates the location of the model with standard
hyperparameter configuration.\label{fig:precision}}
\end{figure}

The disadvantage of MMD is that its absolute value is not interpretable.
Comparatively, precision and recall can provide more insight into how
the model performs. The way we calculate those two values is shown in
Figure \ref{fig:precision}. Essentially, precision indicates how many
generated records are realistic, while recall indicates how many
structures in real samples have been covered by the generative model.
The result is shown in Table \ref{tab:mmd}.

We start by analyzing the result for topological structure. The standard
hyperparameter configuration achieves values of 43.1\% and 78.0\% for
the precision and recall respectively. It is noted that the recall value
is significantly higher than the precision value, meaning that a large
portion of the real dataset can be covered by the model, but the model
will also generate a high portion of molecules that are not realistic.
Comparatively, the precision and recall values for 3D structures are
much higher. The model with standard configuration achieves 81.9\% and
88.3\% precision and recall respectively. This higher value might be
explained by the abstract nature of USRCAT fingerprint. In fact, USRCAT
does not contain explicit information about detailed moleular structure,
but instead focuses on pharmacophore distribution in 3D space. Higher
precision and recall, in this case, might indicate that the model can
better model those 3D pharmacophore features compared with explicit 2D
features.

Precision and recall are two distinct evaluation metrics, and optimizing
those metrics involves multi-objective optimization. Figure
\ref{fig:precision} shows the scatter plot of precision and recall for
topological structures and 3D structures respectively. The Pareto
frontier is colored in blue. The point representing the standard
hyperparameter configuration is indicated using a black arrow. Note that
the standard configuration lies in the Pareto frontier in Figure
\ref{fig:precision} \textbf{c}, and is close to the frontier in Figure
\ref{fig:precision} \textbf{d}.

  \hypertarget{the-validity-of-local-geometries-in-generated-molecules}{%
\subsection{The validity of local geometries in generated
molecules}\label{the-validity-of-local-geometries-in-generated-molecules}}

\begin{figure}[!tbh]
\centering
\includegraphics[width=\linewidth]{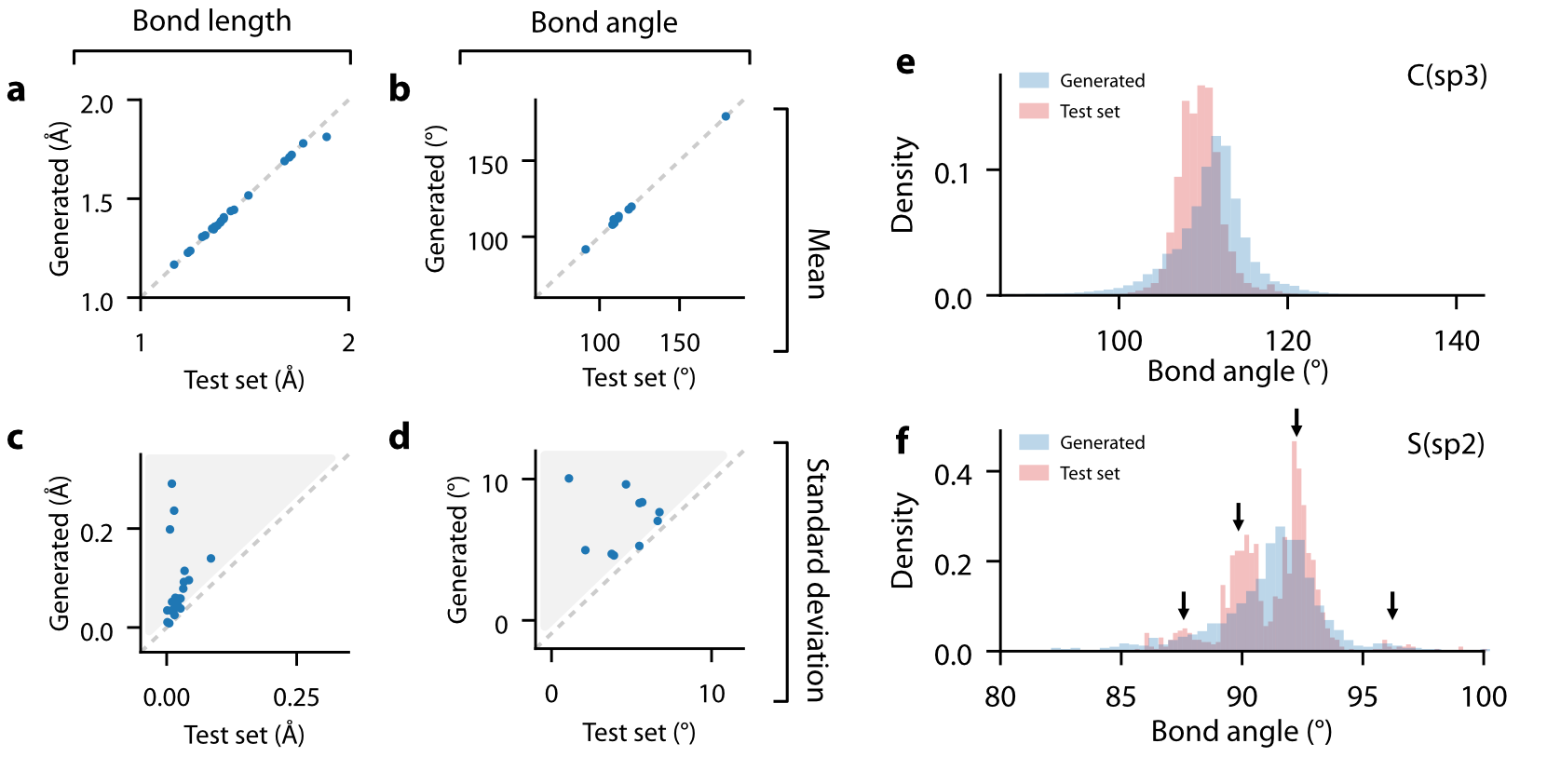}
\caption{Comparing the distribution of bond lengths and bond angles
between generated and test set molecules. \textbf{a-b.} Average bond
lengths (\textbf{a}) and bond angles (\textbf{b}) for each environment
key in generated (y-axis) and test set (x-axis) molecules. \textbf{c-d.}
Standard deviation of bond lengths (\textbf{c}) and bond angles
(\textbf{d}) each environment key in generated (y-axis) and test set
(x-axis) molecules. \textbf{e-f.} The distribution of bond angle in two
atomic environments: \textbf{e.} sp3 hybridized carbon atom; \textbf{f.}
sp2 hybridized sulfur atom.\label{fig:local-struct}}
\end{figure}

The mean and standard deviation for the distribution of bond lengths and
bond angles in different environment are shown in Figure
\ref{fig:local-struct} \textbf{a-d}. The horizontal axis indicates the
statistics for test set molecules, while the vertical axis indicates
that of the generated molecules.

Figure \ref{fig:local-struct} \textbf{a} and \textbf{b} shows that the
model could learn the average bond lengths and angles very well, as all
points in the scatter plot resides closely to the diagonal line (y = x).
For standard deviation (Figure \ref{fig:local-struct} \textbf{c} and
\textbf{d}), it is observed that the points in the scatter plot mostly
reside above the diagonal line, indicating that the model overestimates
the variation in bond lengths and angles. For the standard deviation of
bond length, it can be seen that most points lie close to the diagonal
line, while a small number of points exhibit large deviations. However,
those cases can be easily corrected by manually reducing the standard
deviation during generation.

The mismatch is more significant for the standard deviation of bond
angles. To better understand the problem, we investigate the
distribution of bond angles for two atomic environments (sp3 hybridized
carbon atom and sp2 hybridized sulfur atom), as shown in Figure
\ref{fig:local-struct} \textbf{e} and \textbf{f}. It is found although
there are some deviations in the shape of the distribution (most
significantly for sp2 hybridized sulfur atom, see the black arrows in
Figure \ref{fig:local-struct} \textbf{f}), the model still correctly
predicts the mean and the range of the distribution. Considering that we
have also introduced a mechanism for the model to automatically adapt to
minor errors \ref{data-collection-and-preprocessing}, this mismatch
should not be problematic for the application of the model.

\begin{figure}[!tbh]
\centering
\includegraphics[width=\linewidth]{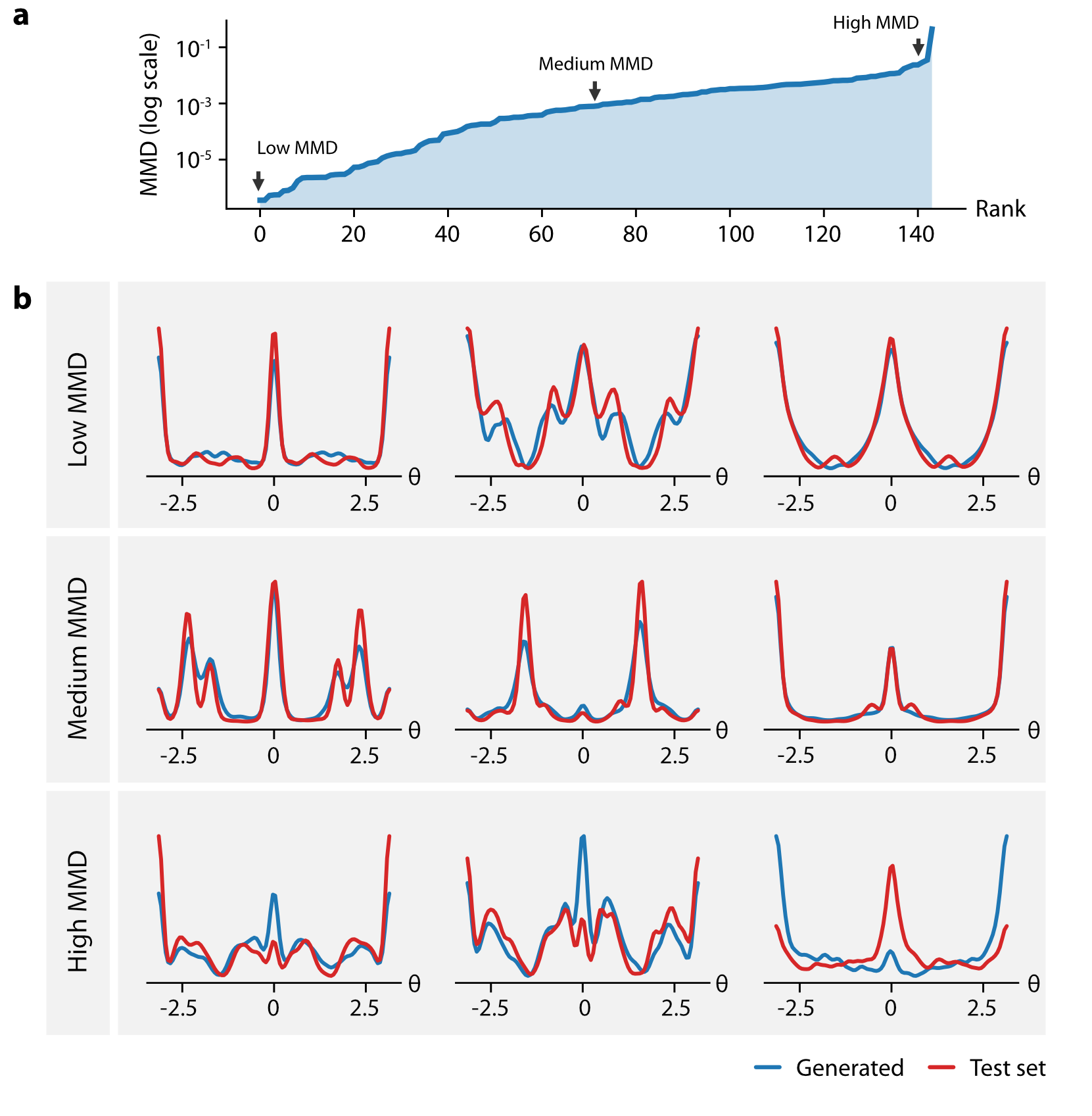}
\caption{Comparing the distribution of torsion angles between generated
and test set molecules. \textbf{a.} The MMD values of the torsion
distribution for each pattern, ranked from lowest to highest.
\textbf{b.} Torsion distributions with the highest, medium and lowest
MMD values. Blue line indicates the generated samples, and red lines
indicates samples in the test set.\label{fig:torsion}}
\end{figure}

Next, we compare the distribution of torsion angles between generated
and test set molecules. The MMD values of the torsion distribution in
different evironment are shown in Figure \ref{fig:torsion} \textbf{a}
(ranking from the lowest MMD to the highest). To make the result more
interpretable, we select three environments with the lowest MMDs, three
environments with medium MMDs, and three environments with the highest
MMDs, and plot the distribution of torsion angles for each case in
Figure \ref{fig:torsion}. The result shows good matches between
generated (blue) and real (red) molecules in 8 out of 9 cases, which
means that the model correctly fits the distribution of torsion angle
for most of the time. A notable exception is the last case, where the
tendency for cis and trans conformation is essentially reverted in the
generated samples.

To summarize, we believe that most of the molecules produced by the
model will have valid local geometry. The results do indicate some
imperfections in the generated structures, but those imperfections are
comparitively small and should have a limited impact on the overall
conformation of the molecule. Also, if desired, those errors can be
easily corrected by including a fast relaxation step using forcefields
such as MMFF94s.

  \hypertarget{quality-of-generated-conformers}{%
\subsection{Quality of generated
conformers}\label{quality-of-generated-conformers}}

\begin{longtable}[]{@{}lll@{}}
\caption{The RMSD values of generated conformers before and after
optimization.\label{tab:rmsd}}\tabularnewline
\toprule
& RMSD & \\
\midrule
\endfirsthead
\toprule
& RMSD & \\
\midrule
\endhead
Methods & mean & std. \\
\hline
Non-random initialization & 0.613 & 0.502 \\
SoftMADE (low noise) & 0.632 & 0.508 \\
No SoftMADE & 0.715 & 0.549 \\
Low input noise & 0.779 & 0.597 \\
Shallow DenseNet & 0.687 & 0.533 \\
Narrow DenseNet & 0.689 & 0.522 \\
Slow decay & 0.623 & 0.503 \\
Fast decay & 0.735 & 0.540 \\
{\bf Standard configuration} & 0.663 & 0.521 \\
\hline
Validation & 0.838 & 0.539 \\
Test & 0.807 & 0.524 \\
\bottomrule
\end{longtable}

\begin{figure}[!tbh]
\centering
\includegraphics[width=\linewidth]{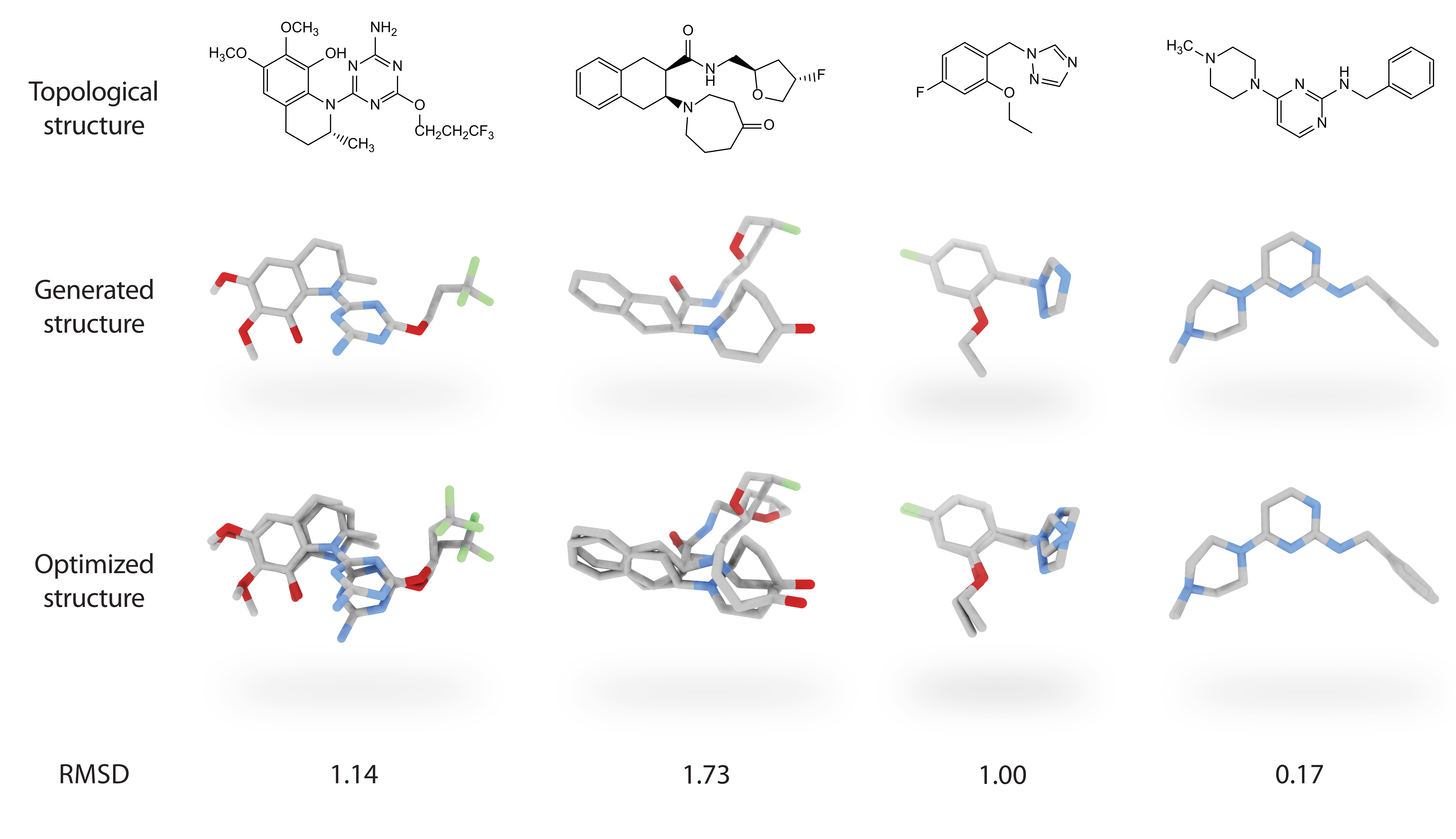}
\caption{The topological structures, generated conformation, optimized
conformation and RMSD value for some generated moelcule from the
model.\label{fig:conf}}
\end{figure}

\begin{figure}[!tbh]
\centering
\includegraphics{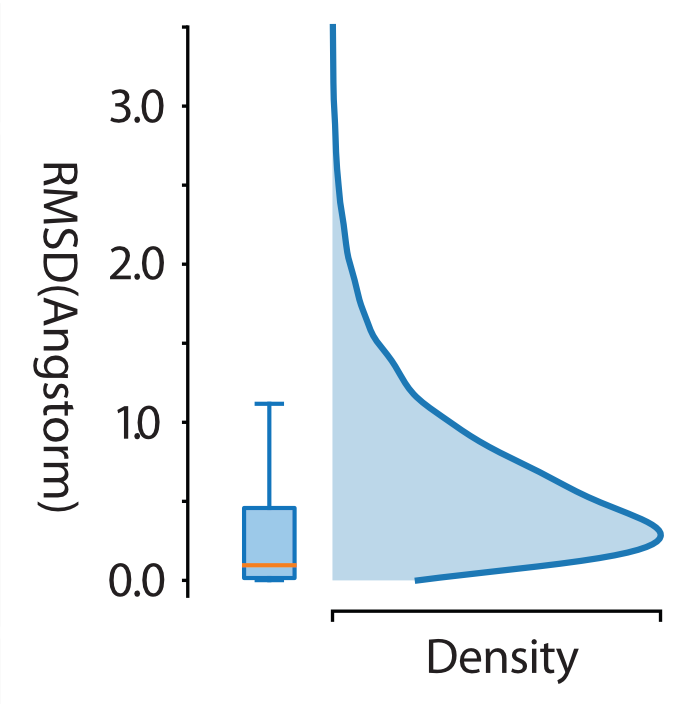}
\caption{The distribution of RMSD values for generated molecules (from
model trained using the standard hyperparameter
configuration)\label{fig:rmsd}}
\end{figure}

We evaluate the quality of generated conformations of molecules by
calculating the RMSD value between the structure before and after the
MMFF94s force field optimization. The average and standard deviation of
RMSD are summarized in Table \ref{tab:rmsd} for each hyperparameter
choice. Figure \ref{fig:conf} shows several randomly generated samples
and the aligned 3D structures before and after optimization. It can be
seen that although the model can not perfectly generate molecules at the
local minima of MMFF94s, the conformation difference before and after
the optimization is relatively small. In fact, the best hyperparameter
set (disable randomized trajectory) is able to achieve an average RMSD
value of 0.613 Å. For other selection of hyperparameters, the value is
generally around 0.7 Å. The variance of the RMSD value is rather large.
Figure \ref{fig:rmsd} shows the distribution of RMSD among generated
molecules, which concentrates in 0-1 Å but have heavy tailing.

To better interpret the RMSD values, we perform the same calculation
using the ETKDG generated conformation of the validation and test set.
ETKDG \cite{Riniker.2015}, which stands for Experimental-Torsion
Distance Geometry with ``basic knowledge'', is a conformation generation
method combining distance geometry with information about torsion
distribution derived from crystallographic data, and is initially
developed as a faster alternative to forcefield based optimization. The
result shows that the RMSD between the optimized and unoptimized
conformation generated using ETKDG is around 0.8 Å (validation set:
0.838 Å, test set: 0.807 Å), which is around 0.2 Å larger than the
best-reported RMSD value. The comparison demonstrated that the
model-generated conformations might be closer to the local minimums of
MMFF94s than that produced by ETKDG. This is an expected outcome since
the model is trained directly with data points sampled using MMFF94s,
while ETKDG is trained without knowing the forcefield. To summarize,
although the model is unable to achieve a perfect result with near-zero
RMSD, it should still able to reach similar or higher performance
compared with ETKDG, which is a currently widely used method for 3d
conformation generation.

  \hypertarget{structure-based-molecule-design-using-l-net-and-mcts}{%
\subsection{Structure-based molecule design using L-Net and
MCTS}\label{structure-based-molecule-design-using-l-net-and-mcts}}

Finally, we combine the L-Net with MCTS to demonstrate its potential in
structure-based molecular design problems. The allosteric site of ABL1
is used as a case study. The structure of ABL1 is shown in Figure
\ref{fig:5mo4} \textbf{a} along with its inhibitor asciminib (PDB ID:
5mo4). The topological structure of asciminib is shown in Figure
\ref{fig:5mo4}. We start with the seed structure containing the
\(CClF_2O-\) group to ensure the inhibition activity against ABL1 and
use MCTS to grow the rest of the structure for high binding affinity
(calculated using smina).

\begin{figure}[!tbh]
\centering
\includegraphics[width=\linewidth]{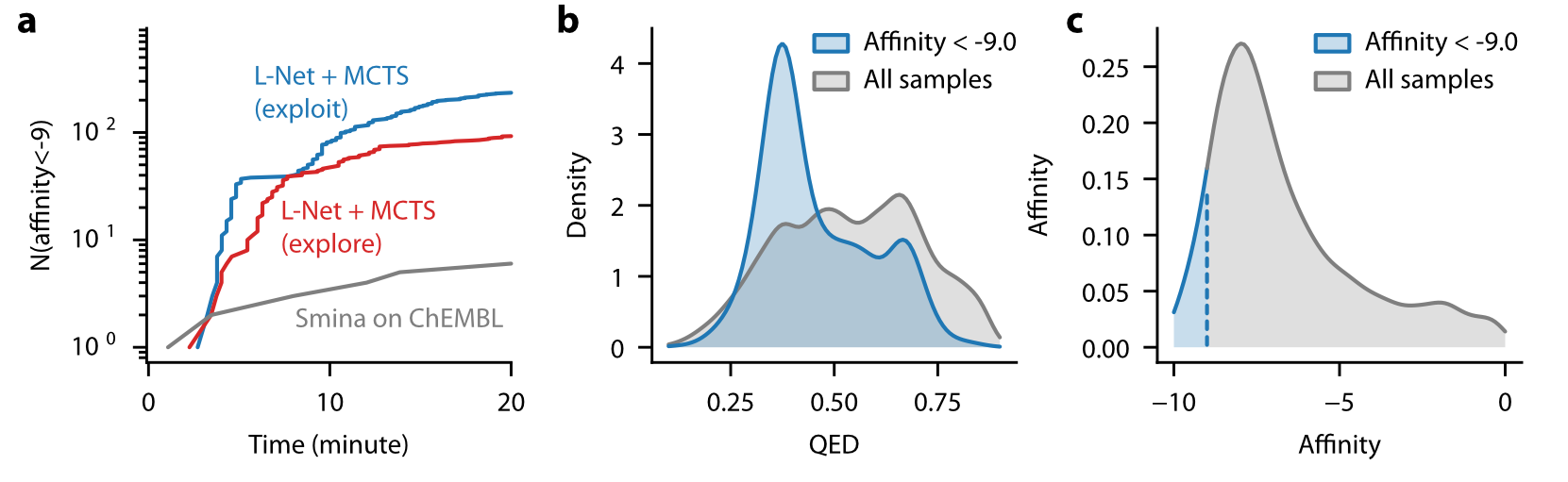}
\caption{\textbf{a.} The speed of model for generating molecules with
high predicted affinity, compared with pure virtual screening with
smina. \textbf{b.} The drug-likeness of all generated samples (grey) and
those with high predicted affinity (blue). \textbf{c.} The distribution
of predicted binding affinity among generated samples. The portion with
high affinity value are highlighted as blue.\label{fig:mcts}}
\end{figure}

We first investigate how fast L-Net and MCTS can generate molecules with
a high predicted binding affinity, defined by a predicted binding
affinity score \textless{} -9.0. Figure \ref{fig:mcts} \textbf{a} shows
how the cumulative number of generated samples with high predicted
binding affinity changes with time. Two hyperparameter sets have been
tried for MCTS, one with a higher exploration factor (the red line), one
with a lower (the blue line). The speed of the model is compared against
that of virtual screening (using smina, performed on the ChEMBL
dataset). We use 1 CPU core and 1 NVIDIA TITAN Xp to run L-Net with
MCTS, and 5 CPU cores to run virtual screening using smina. All models
are run for 20 minutes. The result shows that L-Net combined with MCTS
generates molecules with high predicted binding affinity faster
than pure virtual screening. The speed is faster for MCTS with a lower
exploration factor.

Next, the distribution of drug-likeness (measured with QED) and binding
affinity among generated molecules are examined. We will focus on the
model with a low exploration factor due to its better performance.
Figure \ref{fig:mcts} \textbf{b} shows the distribution of QED among all
generated samples (grey) and those with high predicted affinity (blue).
Note that the overall drug-likeness is lower compared to unconditionally
generated samples (see Figure \ref{fig:prop2d} \textbf{f}), and is even
lower for those with high predicted affinity. This demonstrates the
trade-off the model needs to made between affinity and drug-likeness.
Nevertheless, it can be seen that there is still a significant portion
of highly drug-like molecules (QED \textgreater{} 0.5) among those with
high predicted affinity, providing plenty of candidates for the user to
choose from. The distribution of binding affinity is shown in Figure
\ref{fig:mcts} \textbf{c}, which concentrates around the affinity value
of -7.0. The portion of molecules with high predicted affinity is
colored in blue.

A high-quality subset is subsequently obtained from the generated
molecules. Only molecules with a predicted affinity \textless{} -9.0
will be retained. Those with QED \textless{} 0.5 will be filtered out.
We also filter out results with RMSD \textgreater{} 2.0 before and after
local optimization by smina. The rest of the molecules are manually
inspected and three examples are chosen and are shown in Figure
\ref{fig:5mo4} \textbf{d-f}. We found that most molecules will generate
a benzene ring right after the seed structure. This might have resulted
from the hydrophobicity of that area in the pocket, as we are able to
identify a nearby hydrophobic center using CavityPlus. In comparison,
the amide group in asciminib generally does not occur in the generated
structures. It was previously reported that this structure is involved
in water-mediated H-bonding interaction to the protein. Therefore, the
absence of the amide group might be explained by the lack of water
during smina scoring. The ``top'' portion of the generated structures
generally involves hydrogen bonding with the protein. Note that the
generated structure in Figure \ref{fig:5mo4} \textbf{f} also interacts
with E481, similar to asciminib, but from a different direction. The
predicted binding affinity is also similar for the two molecules (both
have a value of -9.6 after local minimization).

  \hypertarget{conclusion}{%
\section{Conclusion}\label{conclusion}}

In this work, we introduce L-Net, a novel deep generative model for 3D
drug-like molecules. Previous works on this direction have either
focused on structurally simple molecules with limited drug-likeness
\cite{Gebauer.2019,Simm.2020toa,Nesterov.2020}, or is not end-to-end and
requires to combine multiple components and algorithms to
work\cite{Masuda.20201lq,Ragoza.2020}. Comparatively, our proposed
method directly outputs 3D and topological structures of drug-like
molecules, without the need for additional atom placement or bond order
inference. In fact, to our knowledge, we are the first to apply
autoregressive 3D graph generative models to the problem of generating
drug-like molecules with 3D structures. The results show that the model
is capable of generating chemically correct, conformationally valid, and
druglike molecules as output.

We propose a series of techniques to make the model work well for
drug-like molecules. Local coordinate systems are introduced to enforce
rotational equivariance while ensuring the model's expressiveness
(Section \ref{the-embedding-layers}); SoftMADE is used to address the
problems during the fitting of atom position distribution (Section
\ref{decision-making-during-the-append-operation}); Valence and
hybridization state are explicitly considered to improve the quality of
local structures (Section
\ref{decision-making-during-the-append-operation}); Errors (or noise)
are intentionally added to the input so that the model can learn to
correct them (Section \ref{data-collection-and-preprocessing});
Ring-first traversal is introduced to generate ``expert trajectories''
(Section \ref{data-collection-and-preprocessing}). Together, those
methods have contributed significantly to the model performance. We also
developed techniques to reduce the requirement of computational
resources. A hierarchical node clustering method is used to enable
pooling and unpooling on molecular graphs (Section
\ref{pooling-and-unpooling-operations-in-graph-u-net}), reducing the
memory requirement on training and enabling the inclusion of hydrogen
atoms during generation. Optimizations have been applied to accelerate
the generation process (Section
\ref{optimizing-the-speed-of-molecule-generation}).

Due to its end-to-end nature, the L-Net can be conveniently combined
with a variety of techniques to address different design problems in
drug discovery. For example, our method can be directly combined with
reinforcement learning methods, such as policy gradient or MCTS, to
enable objective-oriented molecule generation in 3D space, including
structure-based docking objectives. As a demonstration, we combine L-Net
with MCTS to test its ability in designing potential inhibitors against
ABL1. The result shows that the model is able to generate molecules with
a similar interaction mode and a similar predicted binding affinity
within the pocket as known inhibitors, such as that shown in Figure
\ref{fig:5mo4} \textbf{f}. Based on the results, we believe that the
potential of the proposed method is promising for the application in
structure-based drug discovery.

Although the performance of L-Net is promising, we note that it does
have limitations. Most notably, L-Net is trained using 3D structures
generated from RDKit instead of crystallographic data, which is in
theory more accurate. There are several challenges in using
crystallographic data to train our model. The first one is that the size
of data is more limited compared with automatically generated data,
especially for drug-like molecules. A potential solution to this problem
is to first train on a large artificial dataset generated using RDKit
and then fine-tune the model on a crystallographic dataset. The second
problem is that the crystallographic conformation of small molecules
represents only one conformation (likely the one close to the global
energy minimum). However, in many cases in drug discovery, instead of
requiring the model to strictly generate conformations close to global
energy minimum, we may want a diverse set of conformations to choose
from. A likely solution to this problem is to use molecular dynamics to
augment the crystallographic dataset. We left those explorations to
future works.

  \begin{acknowledgement}
    This work was supported in part by the National Natural Science Foundation 
    of China (Grants 22033001, 21673010 and 21633001), the National Science and Technology 
    Major Project “Key New Drug Creation and Manufacturing Program”, China 
    (Grant 2018ZX09711002), and the Ministry of Science and Technology of China 
    (Grant 2016YFA0502303). We thank Juan Xie for helpful discussions about
    the ABL1 kinase.
\end{acknowledgement}
  \bibliography{main.bib}{}

\end{document}